\documentclass[11pt]{article}
\usepackage[]{graphicx}
\usepackage[]{epsfig}
\newcommand {\eps} {\epsilon}
\newcommand {\sign} {\mbox{sign}}

\newcommand {\la} {\langle}
\newcommand {\ra} {\rangle}

\newcommand {\lam} {\lambda}
\newcommand {\psih} {{\hat \psi}}

\newcommand {\cZ} {{\cal Z}}
\newcommand {\tf} {{\tilde f}}

\def\(({\left(}
\def\)){\right)}
\def\[[{\left[}
\def\]]{\right]}
\newcommand{\nn}{\nonumber}

\newcommand{\be}{\begin{equation}}
\newcommand{\bea}{\begin{eqnarray} \nonumber}
\newcommand{\ee}{\end{equation}}
\newcommand{\eea}{\end{eqnarray}}
\newcommand {\lan} {\langle}
\newcommand {\ran} {\rangle}
 \def\(({\left(}
 \def\)){\right)}
\def\[[{\left[}
\def\]]{\right]}
\def\bi{\bibitem}
\def \form#1 {eq. (\ref{#1}) }
\def \parziale#1#2  {{\partial {#1} \over \partial {#2}}}

\def \cN{{\cal N}}

\def \cQ{{\cal Q}}
\def \cH{{\cal H}}
\def\la{\langle}
\def\ra{\rangle}
\def \Tr {\mbox{Tr}}
\def \ba#1 {\overline{#1}}

\def \eps{\epsilon}

\def\bi{\bibitem}

\newcommand{\douintx}{\int d^{d} x d^{d} y}

\newcommand{\intx}{\int d^{d} x}

\newcommand{\inty}{\int d^{d} y}

\newcommand{\mifun}{\prod_k d \phi_k}
\newcommand{\mifuna}{\prod_{k,a} d \phi^{(a)}_k}
\newcommand{\unoint}{\int \frac{d^{d} q}{(2 \pi)^{d}}}

\newcommand{\bmat}{\begin{mathletters}}
\newcommand{\emat}{\end{mathletters}}
\newcommand{\beq}{\begin{equation}}
\newcommand{\eeq}{\end{equation}}
\newcommand{\bq}{\begin{eqnarray}}
\newcommand{\eq}{\end{eqnarray}}
\newcommand{\beqa}{\begin{eqnarray}}
\newcommand{\eeqa}{\end{eqnarray}}

\def\tr {\mbox{Tr}}
\def\eps{\epsilon}

\def\pia{\phi^{(a)}_i}
\def\psia{\psi^{(a)}}
\def\tpsia{\hat{\psi}^{(a)}}
\def\tpsi{\hat{\psi}^{(1)}}
\def\tchi{\hat{\chi}}
\def\psumma{\sum_{a=1,n} {\left(\hat{\psi}^{(a)}(x)\right)}^2 }
\def\ff{\tilde{f}}
\def\rff{\rho \tilde{f}}
\def\f0{\tilde{f}(0)}
\def\rf0{\rho \tilde{f}(0)}
\def\ep{\epsilon(p)}

\newcommand{\diffe}[1]{\left[ \rff(p-\emph{#1})-\rff(\emph{#1}) \right]}

\begin{document}

\title{Euclidean random matrices: solved and open problems}
%\runningtitle{Euclidean random matrices}
\author{Giorgio Parisi \\
%}\address{
Dipartimento di Fisica, Sezione INFN, SMC and UdRm1 of INFM,\\
Universit\`a di Roma ``La Sapienza'',\\
Piazzale Aldo Moro 2,
I-00185 Rome (Italy)\\
giorgio.parisi@roma1.infn.it}

\date{\empty}

\maketitle
\begin{abstract}
In this paper I will describe some results that have been recently obtained in the study of 
random Euclidean matrices, i.e. matrices that are functions of random points in Euclidean 
space. In the case of {\sl translation invariant} matrices one generically finds a phase transition 
between a {\sl phonon} phase and a {\sl saddle} phase.  
If we apply these considerations
to the study of the Hessian of the Hamiltonian of the particles of a fluid, we find that this
phonon-saddle transition corresponds to the dynamical phase transition in glasses, that
has been studied in the framework of the mode coupling approximation. The Boson peak
observed in glasses at low temperature is a remanent of this transition.
We finally present some recent results obtained with a new approach where one deeply uses some hidden supersymmetric 
properties of  the problem.
\end{abstract}

\section{Introduction}

In the last years many people have worked on the problem of analytically computing the properties of
Euclidean random matrices \cite{MOLTE}-\cite{LOC}.  The problem can be formulated as follows.  

We consider a set of $N$ points ($x_{i}$) that are randomly distributed with some given
distribution.Two extreme examples are: \begin{itemize}
\item
    The $x$'s are random independent points with a flat
probability distribution, with density $\rho$.
\item The $x$'s are one of the many minima of a given
Hamiltonian.  
\end{itemize}
  In the simplest case of the first example, given a function $f(x)$, we consider the $N \times N $
matrix:
\be
M_{i,k}=f(x_{i}-x_{k}) \ . \label{first}
\ee
The problem consists in computing the properties of the eigenvalues and of the 
eigenvectors of $M$.  Of course, for finite $N$ they will depend on the instance of the 
problem (i.e. the actually choice of the $x$'s), however system to system fluctuations 
for  intensive quantities (e.g. the spectral density $g(z)$) disappear when we 
consider the limit $ N \to \infty$ at fixed particle density $\rho$.

The problem is not new; it has been carefully studied in the case where the positions of the
particles are restricted to be on a lattice \cite{loc_rev}.  The case where the particles can stay
in arbitrary positions, that is relevant in the study of fluids, has been less studied, although in
the past many papers have been written on the argument \cite{MOLTE}-\cite{LOC}.  These off-lattice models
present some technical (and also physical) differences with the more studied on-lattice models.

There are many possible physical motivations for studying these models, that may be applied to
electronic levels in amorphous systems, very diluted impurities, spectrum of vibrations in glasses
(my interest in the subject comes from this last field).  

I will concentrate in these lectures on
the spectral density and on the Green functions, trying to obtaining both the qualitative features
of these quantities and to present reasonable accurate computations (when possible).  This task is
not trivial because there is no evident case that can be used as starting point for doing a
perturbation expansion.  Our construction can be considered as a form of a mean field theory (taking
care of some corrections to it): more sophisticated problems, like localization or corrections to
the naive critical exponents will be only marginally discussed in these notes.

A certain number of variations to equation (\ref{first}) are interesting.  For example we could
consider the case where we add a fluctuating term on the diagonal:
\be
M_{i,k}=\delta_{i,k} \sum_{j}f(x_{i}-x_{j}) -f(x_{i}-x_{k}) \ . \label{TT}
\ee
This fluctuating diagonal term has been constructed in such a way that
\be
\sum_{k}M_{i,k}=0 \ .
\ee
Therefore the diagonal and the off-diagonal matrix elements are correlated.

In general it may be convenient to associate a quadratic form  to the matrix $M$: the quadratic form
($\cH[\phi]$) is defined as:
\be
\cH[\phi]= \frac12 \sum_{i,k}\phi_{i}\phi_{k}M_{i,k}\ . 
\ee
In the first case we considered, eq. (\ref{first}), we have that:
\be
\cH[\phi]= \sum_{i,k}f(x_{i}-x_{k})  \phi_{i}\phi_{k}\ .\label{FIRST}
\ee
In this second case, eq. (\ref{TT}), the associated quadratic form is given by
\be
\cH[\phi]=\sum_{i,k}\phi_{i}\phi_{k}M_{i,k}=
\frac12 \sum_{i,k}f(x_{i}-x_{k}) ( \phi_{i}-\phi_{k})^{2} \label{SECOND} \ .
\ee
Here the matrix $M$ is non-negative if the function $f$ is non-negative.  The matrix $M$
has always a zero eigenvalue as consequence of the invariance of the quadratic form under the
symmetry $\phi_{i} \to \phi_{i} +\lambda$; the presence of this symmetry has deep consequences on
the properties of the spectrum of $M$ and, as we shall see later, {\sl phonons} are present.  Many
of the tricky points in the analytic study are connected to the preserving and using this symmetry
in the computations.

In the same spirit we can consider a two-body potential $V(x)$ and we can introduce the 
Hamiltonian 
\be
H[x]=\frac12 \sum_{i,k}V(x_{i}-x_{k}) \ . \label{HAMI}
\ee
We can construct the $3N \times 3N $ Hessian matrix
\be
M_{i,k}= {\partial^{2}H\over \partial x_{i} x_{k}}=
\delta_{i,k} \sum_{j}V''(x_{i}-x_{j}) -V''(x_{i}-x_{k}) \ ,\label{HESSIAN}
\ee
where for simplicity we have not indicated space indices.
Also here we are interested in the computation of the spectrum of $M$.  The translational 
invariance of the Hamiltonian implies that the associated quadratic form is invariant 
under the symmetry $\phi_{i} \to \phi_{i} +\lambda$ and a phonon-like behavior may be 
expected.

This tensorial case, especially when the distribution of the $x$'s is related to the potential 
$V$, is the most interesting from the physical point of view (especially for its 
mobility edge \cite{loc_rev,LOC}).  Here we stick to the much simpler question of the computation of 
the spectrum of $M$.  We shall develop a field theory for this problem, check it at high and low 
densities, and use a Hartree type method.

Our aim is to get both a qualitative understanding of the main properties of the spectrum and of the
eigenvalues and a quantitate, as accurate as possible, analytic evaluation of these properties.
Quantitative accurate results are also needed because the computation of the spectral density is a
crucial step in the microscopic computation of the thermodynamic and of the dynamic properties of
glass forming systems.  In some sense this approach can be considered as an alternative route to
obtain mode-coupling like results \cite{TDYN}, the main difference being that in the mode coupling
approach one uses a coarse grained approach and the hydrodynamical equations, while here we take a
fully microscopic point of view.

Many variations can be done on the distribution of points, each one has its distinctive flavor:
\begin{itemize}
	 \item The distribution of the points $x$ is flat and the points  are uncorrelated: they are
	 uniformly distributed in a cube of size $L$ and their number is $N=\rho L^{3}$. Here, as in the
	 following cases, we are interested in the thermodynamic limit where both $L$ and $N$ go to
	 infinity at fixed $\rho$.
	 \item The points $x$ are distributed with a distribution that is proportional to $ \exp(- \beta
	 H[x]) $ where $H[x]$ is a given function.
	 \item The points $x$ are one of the many solutions of the equation $\partial H/\partial x_{i}=0$.  This
	 last problem may be generalized by weighting each stationary point of $H$ with an
	 appropriate weight.
\end{itemize}

The last two cases are particularly interesting when 
\be
M_{i,k}= {\partial^{2}H\over \partial x_{i} x_{k}} \ ,
\ee
and consequently
\be
\cH[\phi]=\frac12 \sum_{i,k}\phi_{i}\phi_{k}{\partial^{2}H\over \partial x_{i} x_{k}} \ .
\ee
If this happens the distribution of points and the matrix are related and  the theoretical challenge is to use
this relation in an effective way.

In the second section of these lectures, after the introduction we will present the basic definition
of the spectrum, Green functions, structure functions and related quantities. In the third section
we will discuss the physical motivation for this study and we will present a first introduction to
the replica method. In the fourth section we will give a few example how a field theory can be
constructed in such way that it describes the properties of randomly distributed points. In the
fifth section we will discuss in details the simples possible model for random Euclidean matrices,
presenting both the analytic approach and some numerical simulations. In the sixth section we shall 
present a similar analysis in a more complex and more physically relevant case, where phonons are
present due to translational invariance. Finally, in the last section we present some recent results
that have been obtained in the case of correlated points.

\section{Basic definitions}
The basic object that we will calculate are the matrix element of the resolvent
\be 
G[x](z)_{i,j}=\((\frac{1}{z - M[x]}\))_{ij} \ ,
\ee
(we use the notation $M[x]$ in order to stress that the matrix $M$ depends on all the points $[x]$.  
Sometimes we will suppress the specification ``$[x]$'', where it is obvious.  We can define the sample 
dependent Green function
\bea
G[x](y_{1},y_{2},z)=\sum_{i,j}\delta(y_{1}-x_{i})\delta(y_{2}-x_{j})G[x](z)_{i,j} =\\
\sum_{i,j}\delta(y_{1}-x_{i})\delta(y_{2}-x_{j})\((\frac{1}{z - M[x]}\))_{ij} \ .
\eea
The quantity that we are interested to compute analytically are the sample averages of the Green function, 
i.e. 
\be
\overline G(y_{1},y_{2},z)=\overline{G[x](y_{1},y_{2},z)}\,,
\ee
where the overline denotes the average over the random position of the points.

If the problem is translational invariant,  after the average on the positions of the points $[x]$ we have
\be
\overline G(y_{1},y_{2},z)=\overline  G(y_{1}-y_{2},z) \ ,
\ee
where the function $\overline  G(x,z)$ is smooth apart from a delta function at $x=0$.
It is convenient to consider the Fourier transform of $\overline G(x,z)$, i.e. 
\bea
G(p,z) = {1 \over N}
\overline{\sum_{ij} e^{i p(x_i-x_j)} \left[\frac{1}{z - M}\right]_{ij} } =\\
{1 \over N}\overline{\int dx dy e^{i p(x-y)} G[x](x,y,z) }=
\int dx e^{i p x} \tilde G(x,z)
\ ,
\label{GDIPI}
\eea
The computation of  the function $G(p,z)$ will  one of the main goals of these lectures.
It is convenient to introduce the so called dynamical structure factor (that in many case can be observed 
experimentally) that here is defined as \footnote{
We shall often make use of the distribution identity \hbox{$1/(x+i0^{+})=P(1/x)-{\mathrm i}\pi\delta(x)$} 
where \hbox{$P(1/x)$} denotes the principal part, i.e. \hbox{$1/(x+i0^{+})+1/(x+i0^{-})=2P(1/x)$}.}:
\begin{equation}
S_e(p,E)=-\frac{1}{\pi}\lim_{\eta\to 0^+} G(p,E+{\mathrm i}\eta) \  .
\end{equation}
One must remember that the structure function is usually defined using as variable $\omega=\sqrt{E}$:
\be
S(p,\omega)=2\omega S_e(p,\omega^2)\,,
\ee
The resolvent will also give us access to the density of states of
the system:
\begin{eqnarray}\nn
g_e(E)&=&
-\frac{1}{N\pi}\lim_{\eta\to 0^+}\overline{\sum_{i=1}^N 
\left[\frac{1}{E+{\mathrm i}\eta - M}\right]_{ii}}\,,\\
&=&-\frac{1}{\pi}\lim_{\eta\to 0^+}\lim_{p\to\infty}G(p,E+{\mathrm i}\eta)\,.
\label{DOF}
\end{eqnarray}

\section{Physical motivations}

There are many physical motivations for studying these problems, apart from the purely 
mathematical interest. Of course different applications will lead to different forms of the problems.
In this notes I will concentrate my attention on the simplest models, skipping the complications (that 
are not well understood) of more complex and interesting models.
\subsection{Impurities}
We can consider a model where there are some impurities in a solid that are localized at the points $x$'s.
There are many physical implementation of the same model; here we only present two cases:
\begin{itemize}
	 \item There may be electrons on the impurities and the amplitude for hopping from an impurity ($i$) to an 
	 other impurity ($k$) is proportional to $f(x_{i}-x_{k})$. The electron density is low, so that the 
	 electron-electron interaction can be neglected.
	 \item There is a local variable $\phi_{i}$ associated to the impurity (e.g. the magnetization) and the
	 effective Hamiltonian at small magnetizations is given by eq.(\ref{FIRST}).
\end{itemize}

In both cases it is clear that the physical behavior is related to the properties of the matrix $M$
that is of the form discussed in the introduction.  If the concentration of the impurities is small
(and the impurities are embedded in an amorphous system) the positions of the impurities may be
considered randomly distributed in space.

\subsection{Random walks}

Let us assume that there are random points in the space and that there is a particle that hops from one 
point to the other with probability per unit time given by $f(x_{i}-x_{k})$.

The evolution equation for the probabilities of finding a particle at the point $i$ is given by
\be
{dP_{i}\over dt}= \sum_{k}f(x_{i}-x_{k})P_{k} - P_{i}\sum_{k}f(x_{i}-x_{k})=
-\sum_{k}M_{i,k}P_{k} \,,
\ee
where the matrix $M$ is given by eq. (\ref{TT}).

Let us call $P(x,t)$ the probability that a particle starting from a random point at time 0 arrives at 
at time $t$. It is evident that after averaging of the different systems (or in a large system after 
averaging over the random starting points) we have that
\be
\int_{0}^{\infty}dt \exp(-tz) \int dx \exp(i px) P(x,t) =-G(p,-z)\,,
\ee
so that the function $P(x,t)$ can be reconstructed from the knowledge of the function $G(p,z)$
defined in eq. (\ref{GDIPI}).

\subsection{Instantaneous  modes}
We suppose that the points $x$'s are distributed according to the probability distribution
\be
\exp(-\beta H[x])\ ,
\ee
where $H[x]$ is for example a two body interaction given by eq.(\ref{HAMI}) and $\beta=1/(kT)$.

The eigenvectors of the Hessian of the Hamiltonian (see eq.(\ref{HESSIAN})) are called instantaneous normal modes
(INN).  The behavior of the INN at low temperature, especially for systems that have a glass
transition \cite{Zim,Ell}, is very interesting \cite{KEYES}.  In particular there have been many
conjectures that relate the properties of the spectrum of INN to the dynamic of the system when
approaching the glass transition.

However it has been realized in these years that much more interesting quantities are the instantaneous 
saddle modes (ISN) \cite{C,ab,CaGiPa,P12}.  Given an equilibrium configuration $x_{e}$ at temperature $T$ 
one defined the inherent saddle $x_{s}$ corresponding to that configuration $x_{e}$ in the following way.  
We consider all the saddles of the Hamiltonian, i.e. all the solutions of the equations
\be
{\partial H\over\partial x_{i}}=0\,.
\ee
The inherent saddle $x_{s}$ is the nearest saddle to the configuration $x_{e}$. The Boltzmann
distribution at temperature $T$ induces a probability distribution on the space of all saddles. The 
physical interesting quantities are the properties of the Hessian around these saddles. It turns out that 
they have a very interesting behavior near the glass transition in the case of glass forming materials.

It is clear that the computation of the spectrum in these case is much more complex: we have both to be 
able to compute the correlations among the points in this non-trivial case and to take care of the 
(often only partially known) correlations.

A possible variation on the same theme consists in consider the ensemble of all saddles of given energy
$E=H[x]$: there are indications that also in this case there may be a phase transition in the properties of the
eigenvalues and eigenvectors when $E$ changes and this phase transition may be correlated to some
experimentally measured properties of glassy systems, i.e. the Boson Peak, as we will see later.

\section{Field theory}\label{FIELDTH}

\subsection{Replicated Field Theory}

The resolvent $G(p,z)$ defined in eq.(\ref{GDIPI}) can be written as a
propagator of a Gaussian theory:
\bea
{\cal Z}[J]& \equiv& \int{\mifun  \exp\((-{1 \over 2} \sum_{lm}
\phi_l \left( z-M \right)_{lm} \phi_m + \sum_l \phi_l J_l \))}\,,\\
\label{DISCREFUNZIO}
 G(p,z)&=& {1 \over N} \sum_{ij} \frac{\delta^2}{\delta J_i \delta
J_j} \left. \overline{\exp(i p(x_i-x_j)) \ln {\cal Z}[J] \: }
\right|_{J=0}\, ,
\label{GDIPIGAU}
\eea where the overline denotes the average over the distribution $P[x]$ of the positions of the particles.
We have been sloppy in writing the Gaussian integrals (that in general are not convergent): one should suppose
that the $\phi$ integrals go from $-a \infty $ to $+a\infty $ where $a^{2}=-i $ and $z$ has a positive (albeit
infinitesimal) imaginary part in order to make the integrals convergent \cite{loc_rev,PARISILOC}.  This choice
of the contour is crucial for obtaining the non compact symmetry group for localization \cite{PARISILOC}, but
it is not important for the density of the states.

In other words, neglecting the imaginary factors, we can introduce a probability distribution over the field
$\phi_{i}$,
\be
 P[\phi] \propto \ \mifun \exp\((-{1 \over 2} \sum_{lm} \phi_l \left( z-M \right)_{lm} \phi_m \)) \ .
\ee
If we denote by $\lan \cdot \ran$ the average with respect to this probability (please notice that here
everything depends on $[x]$ also if this dependence is not explicitly written) we obtain the simple result:
\be
\lan \phi_{i}\phi_{k}\ran =\ G_{i,k} \ .
\ee

A problem with this  approach is that the normalization factor of the probability depends on $[x]$
and it is not simple to get the $x$-averages of the expectation values.  Replicas are very useful in this
case.  They are an extremely efficient way to put the dirty under a simple carpet (on could also use the
supersymmetric approach of Efetov where one puts the same dirty under a more complex carpet \cite{loc_rev}).

For example the logarithm in eq.(\ref{GDIPIGAU}) is best dealt with using the replica trick
\cite{MOLTE,MPZ,CaGiPa}:
\be
\ln{\cal Z}[J] = \lim_{n \to 0}\frac{1}{n} \left( {\cal Z}^n[J] -1 \right)\,.
\ee
The resolvent can then be computed from the $n$-th power of ${\cal Z}$, that
can be written using $n$ {\em replicas}, $\phi^{(a)}_i$ ($a=1,2,\ldots,n$) of the
Gaussian variables of eq.(\ref{DISCREFUNZIO}):

\begin{eqnarray}\nn
G(p,z) = \lim_{n \to 0}  G^{(n)}(p,z) \ , 
\\
 N G^{(n)}(p,z)=  \\
\sum_{ij}   
\overline{\exp(i p(x_i-x_j)) \int \mifuna\, 
\phi^{(1)}_i\phi^{(1)}_j\, 
\exp\((-{1 \over 2} \sum_{l,m,c}
\phi^{(c)}_l \left( z-M \right)_{lm} \phi^{(c)}_m\))}\, . \label{GDISCRE}\nn
\end{eqnarray}
Indeed 
\begin{eqnarray}\nn
\int \mifuna\, 
\phi^{(a)}_i\phi^{(b)}_j\, 
\exp\((-{1 \over 2} \sum_{l,m,c}
\phi^{(c)}_l \left( z-M \right)_{lm} \phi^{(c)}_m\))\propto \delta_{a,b}=\\
\[[{1\over z-M}\]]_{i,j} \det(z-M)^{-n/2}
\end{eqnarray}
and the physically interesting case is obtained only when $n=0$.

In this way one obtains a $O(n)$ symmetric field theory.  It well known that an $O(n)$ symmetric theory is
equivalent to an other theory invariant under the $O(n+2|1)$ symmetry (\cite{RFIM}), where this group is
defined as the one that leaves invariant the quantity
\be
\sum_{a=1,n+2}(\phi^{a})^{2}+\bar \psi \psi \,,
\ee
where $\psi$ is a Fermionic field. Those who hate analytic continuations, can us the group 
$O(2|1)$ at the place of the more fancy $O(0)$ that is defined only as the analytic continuation to $n=0$ of
the more familiar $O(n)$ groups.

Up to this point everything is quite general.  We still have to do the average over the random positions.
This can be done in an explicit or compact way in some of the previous mentioned cases and one obtains a
field theory that can be studied using the appropriate approximations.  Before doing this it is may be
convenient to recall a simpler result, i.e. how to write a field theory for the partition function of a fluid.

\subsection{The partition function of a fluid}

Let us consider a system with $N$ (variable) particles characterized by a classical Hamiltonian
$H_{N}[x]$ where the variable $x$ denote the $N$ positions.  In the simplest case the particles can
move only in a finite dimensional region (a box $B$ of volume $V_{B}$) and they have only two body
interactions:
\be
H_{N}[x]=\frac12 \sum_{i,k}V(x_{i}-x_{k}) \ .
\ee
At given $\beta$ the canonical partition function can be written as
\be
\cQ_{N}=\int_{B}dx_{1}\ldots dx_{N} \exp( -\beta H_{N}[x]) \ ,
\ee
while the gran-canonical partition function is given by 
\be
\cQ(\zeta)=\sum_{N}{\zeta^{N}\over N!}\cQ_{N} \ ,
\ee
where $\zeta$ is the fugacity.

We aim now to find out a field theoretical representation of the $\cQ(\zeta)$. This can be formally 
done  \cite{POLIA} 
(neglecting the problems related to the convergence of Gaussian integrals) by writing
\bea
\exp\(( - \frac{\beta}2 \sum_{i,k}V(x_{i}-x_{k})\))=\\
\cN^{-1} 
 \int d\sigma \exp \(( \frac1{2\beta} \int dx dy V^{-1}(x-y) \sigma(x) \sigma(y) +\sum_{i}\sigma(x_{i})\)) \ ,
 \label{Q1}
\eea
where $\cN$ is an appropriate normalization factor such that 
\begin{eqnarray}\nn
\cN^{-1} \int d\sigma \exp \(( \frac1{2\beta}  \int dx dy V^{-1}(x-y) \sigma(x) \sigma(y) +\int 
J(x)\sigma(x)\))= \\
\exp \(( - \int dx dy \beta V(x-y)J(x)J(y)\)) \ .
\end{eqnarray}
The reader should notice that $V^{-1}(x-y)$ is formally defined by the relation
\be
\int dz V^{-1}(x-z)V(z-y)=\delta(x-y) \ .
\ee
The relation eq.(\ref{Q1}) trivially follows  from the previous equations if we put
\be
J(x)=\sum_{i}\delta(x-x_{i}) \ .
\ee
It is convenient to us the notation  
\begin{equation}
d\mu[\sigma]\equiv \cN^{-1} d\sigma \exp \(( \frac1{2\beta}  \int dx dy V^{-1}(x-y) \sigma(x) 
\sigma(y)\)) \ .
\end{equation}
With this notation  we find
\be
\cQ_{N}=\int d\mu[\sigma] \((\int dx \exp(\sigma(x)\))^{N} \ .
\ee
We finally get
\be
\cQ(\zeta)=\int d\mu[\sigma] \exp \((\zeta \int dx \exp(\sigma(x))\)) \ .
\ee

People addict with field theory can use this compact representation in order to derive the virial 
expansion, i.e. the expansion of the partition in powers of the fugacity at fixed $\beta$.

The same result can be obtained in a more straightforward way \cite{MIG} by using a functional integral
representation for the delta function:
\bea
\exp\((- \frac{\beta}2 \sum_{i,k}V(x_{i}-x_{k})\))= \\
\int d[\rho]\delta_{F}\[[\rho(x)-\sum_{i=1,N}\delta(x-x_{i})\]]
\exp\((-\frac{\beta}2  \int dx dy \rho(x)\rho(y) V(x-y)\))\ ,
\eea
where $\delta_{F}$ stands for a functional Dirac delta:
\be
 \delta_{F}[f]=\int d\sigma \exp \((i\int dx f(x) \lambda(x) \))\ .
\ee
We thus find 
\begin{eqnarray}
\cQ_{N}= 
\int dx_{1} dx_{N}  \int d[\rho] d[\lambda]\\
 \exp\((-\frac{\beta}2 \int dx dy \rho(x)\rho(y) V(x-y) +i \lambda(x) \rho(x) -i \sum_{i=1,N} 
 \lambda(x_{i})\)) \nn \\
 = \int d[\rho] d[\lambda]
 \exp\((-\frac{\beta}2 \int dx dy \rho(x)\rho(y) V(x-y)+i \lambda(x) \rho(x)\)) \nn \\ 
 \((\int dx \exp(i \lambda(x)\))^{N} \ .
 \nn
 \eea

At the end of the day we get
\be
\cQ(\zeta)= \int d[\lambda]
 \exp\((-\frac{\beta}2 \int dx dy \lambda(x)\lambda(y) V^{-1}(x-y) +\zeta \int dx \exp(i \lambda(x)\)) \ .
 \ee
where, by doing the Gaussian integral over the $\rho$, we have recover  the previous  
formula eq.(\ref{Q1}), if we set $i\lambda(x)=\sigma(x)$.

In order to compute the density and its correlations it is convenient to introduce an external field
\be
\sum_{i}U(x_{i}) \ . 
\ee
This field is useful because its derivatives of the partition function give the density correlations. As 
before the partition function is given by 
\be
\cQ(\zeta|U)=\int d\mu[\sigma] \exp \((\zeta \int dx \exp(\sigma(x)+\beta U(x))\)) \ .
\ee

In this way one finds that the density and the two particle correlations are given by:
\bea
\rho=\lan \sum_{i}\delta(x-x_{i})\ran =\zeta \lan  \exp(\sigma(0))\ran \\
C(x)=\lan\sum_{i,k}\delta(x_{i}-0)\delta(x_{k}-x)\ran= \zeta^{2}\lan  \exp(\sigma(0)+\sigma(x))\ran \ .
\eea

In the low density limit (i.e. $\zeta$ near to zero) one finds
\bea
\rho=\zeta\exp(-\beta V(0)) \,,\\
C(x)=  \rho^{2}  \exp(-\beta V(x))\ .
\eea
that is the starting point of the virial expansion.

\section{The simplest case}
In this section we shall be concerned with simple case for random Euclidean matrices, where the positions
of the particles are completely random, chosen with a probability $P(x)=\rho/V_{B}$.

We will consider here the simplest case, where 
\be
M_{i,k}=f(x_{i}-x_{k}) \label{pb}
\ee
and $f(x)$ is bounded, fast 
decreasing function at infinity (e.g. $\exp(-x^{2}/2)$).  

We shall study the field-theory perturbatively 
in the inverse density of particles, $1/\rho$.  The zero$^{th}$ order of this expansion (to be calculated 
in subsection~\ref{HIGHDEN}) corresponds to the limit of infinite density, where the system is 
equivalent to an elastic medium.  In this limit the resolvent (neglecting $p$ independent terms) is extremely simple:
\be
G(p,z)=\frac{1}{z-\rho\ff(p)}\, . \label{ZEROTH}
\ee
In the above expression $\ff(p)$ is the Fourier transform of the function $f$, that due to its spherical 
symmetry, is a function of only $(p^2)$ .  We see that the dynamical structure function has two delta 
functions at frequencies \hbox{$\omega=\pm\sqrt{\rho\ff(p)}$}.

It is then clear that Eq.(\ref{ZEROTH}) represents the undamped propagation of plane waves in our harmonic
medium, with a dispersion relation controlled by the function $\ff$ .  The high order corrections to
Eq.(\ref{ZEROTH}) (that vanishes as $1/\rho$) will be calculated later.  They take
the form of a complex self-energy, $\Sigma(p,z)$, yielding
\begin{eqnarray}
G(p,z) &=&  \frac{1}{z-\ep -\Sigma(p,z)}
\label{CORRETTO}\,,\\
S_E(p,E)&=&  -\frac{1}{\pi} \frac{Im \Sigma(p,E)}
{\left(E-\ep -Re \Sigma(p,E)\right)^2 +\left(Im \Sigma(p,E)\right)^2 }
\label{ALLARGAMENTO}\,.
\end{eqnarray}
The dynamical structure factor,
is no longer a delta function, but close to its maxima it has a Lorentzian shape.  From
eq.  (\ref{ALLARGAMENTO}) we see that the real part of the self-energy renormalizes the dispersion relation.
The width of the spectral line is instead controlled by the imaginary part of the self-energy.

subsection{Field theory}
Let us first consider the spectrum: it  can be computed from the imaginary part of the trace of the resolvent:
\be
R(z)={1 \over N}\overline{ \Tr {1 \over z-M}} \ ,
\label{res_def}
\ee
where the overline denotes the average over the positions $x_i$.

It is possible to compute the resolvent from a field theory written using a replica
approach. We shall compute $\Xi_N \equiv \overline{ \det (z-M) ^{-n/2}} $,
and deduce from it the resolvent by
using the replica limit $n \to 0$.

It is easy to show that
one can write $\Xi_N$ as a partition function
over replicated fields $\phi_i^a,$ where $i\in\{ 1...N\} , \ a\in \{1...n\} $:
\bea
\Xi_N=
\int \prod_{i=1}^N {dx_i \over V} \int \prod_{i=1}^N \prod_{a=1}^n
d \phi_i^a \\
\exp\((-{z \over 2} \sum_{i,a}\((\phi_i^a\))^2+ {1 \over 2} \sum_{i,j,a}
f(x_i-x_j) \phi_i^a \phi_j^a\)) \ .
\eea
In order to simplify the previous equations let us introduce the Bosonic \footnote{In spite of a long
tradition, here the fields $\psi$ are Bosons, not Fermions.  Fermionic fields will appear only in the last
section of these note.} fields $\psi_a(x)=\sum_{i=1}^N \phi_i^a \delta(x-x_i)$ together with their respective Lagrange
multiplier fields $\hat \psi_a(x)$.  More precisely we introduce a functional delta function:
\bea
\delta_{F}\(( \psi_a(x)-\sum_{i=1}^N 
\phi_i^a \delta(x-x_i) \))= \\
\int d[\hat \psi_{a}] \exp\((i \int dx \hat \psi_{a}(x)\((\psi_a(x)-\sum_{i=1}^N 
\phi_i^a \delta(x-x_i)\)) \))\ .
\eea
One 
can integrate out the $\phi$ variables, leading to the following field theory for $\Xi_N$, where we have 
neglected all the $z$ independent factors that disappear when $n=0$:
\be
\Xi_N= \int D[\psi_a,\hat \psi_a] A^N
\exp\((S_0\))
\ee
where
\bea
S_0&=&i \sum_a \int dx\  \psih_a(x) \psi_a(x)
+ {1 \over 2} \sum_a
\int dx dy \ \psi^a(x) f(x-y) \psi^a(y)  \ ,\\
A&=& 
\exp\[[-{1 \over 2 z}
 \sum_{a} \psih^a(x)^2 \]] \ .\label{GRAN}
\eea

It is convenient to go to a grand canonical formulation for the disorder: we
consider an ensemble of samples with varying number of ($N$), and compute the grand canonical
partition function
${\cZ}(\zeta) \equiv \sum_{N=0}^\infty \Xi_N \zeta^N /N!$
that is equal to:
\be
\cZ= \int D[\psi_a,\hat \psi_a]  \exp\((S_0 + \zeta A\)) \ \  ; \  \
\ .
\label{cZ1}
\ee

In the $n \to $ limit, one finds that $\rho=\zeta$ .  The definition of the field $\psi$ implies
that the correlation is given (at $x\ne 0)$ by:
\be
\overline  G(x,z) \equiv \sum_{i,k}\delta(x-x_{i})\delta(x_{j})\(({1 \over z-M}\))_{i,k}
=\lan \psi_{a}(x) \psi_{a}(0)\ran
\ee
A simple computation (taking care also of the contribution in $x=0$) gives
\be
G(p,z) =   \frac{1}{\rho \ff (p)} - \lim_{n \to 0} \frac{1}{\rho \ff^2 (p)} \douintx 
\: \exp(i 
p(x-y)) \: \langle  \tpsi (x) \tpsi (y) \rangle  \, .
\label{GFINALE}
\ee
so that the average Green function can be recovered from the knowledge of  the $ \psih$ propagator.
 
Notice that we can also integrate out the $\psi$ field thus replacing
$S_0$ by $S_0'$, where
\be
S_0'= {1 \over 2} \sum_a
\int dx dy \ \psih^a(x) f^{-1}(x-y) \psih^a(y),
\label{cZ2}
\ee
where $f^{-1}$ is the integral operator that is the inverse of $f$.

The expression (\ref{cZ1}) is our basic field theory representation.  We shall denote by brackets the 
expectation value of any observable with the action $S_0+S_1$.  As usual with the replica method we have 
traded the disorder for an interacting replicated system.

It may be convenient to present a different derivation of the previous result: we write
\bea
\exp\((-{z \over 2} \sum_{i,a} (\phi_i^a)^2+ {1 \over 2} \sum_{i,j,a}
f(x_i-x_j) \phi_i^a \phi_j^a\))=\\
\int d[\omega]\exp \((-{1 \over 2}\sum_{a} \int dx dy\; \omega(x)_{a}\omega(y)_{a} f^{-1}(x-y)+\sum_{i,a}\((
 \omega(x_{i})_{a}\phi(x_{i})_i^a -\frac{z}2(\phi_i^a)^2 \)) \)) \ .
 \eea
If we collect all the terms  at a given site $i$, we get
\be
 \int \prod_a d\phi_i^a\exp \((\sum_{a}\((
 \omega(x_{i})_{a}\phi(x_{i})_i^a -\frac{z}2(\phi_i^a)^2 \)) \)) \ .
\ee
The $\phi$ integrals are Gaussian and they can be done: for $n=0$ one remains with 
\be
\exp\((\sum_{a} \omega(x_{i})^{2}_{a}/z \))
\ee
and one recovers easily the previous result eq.(\ref{GRAN}).

The basic properties of the field theory are related to the properties of the original problem in a
straightforward way.  We have seen that the average number of particles is related to $\zeta$ through $N=\zeta
V \la A \ra$, so that one gets $\zeta=\rho$ in the $n \to 0$ limit because $\la A \ra=1$.  From the
generalized partition function $\cZ$, one can get the resolvent $R(z)$ through:
\be
R(z)=- \lim_{n \to 0} {2 \over n N} {\partial \log \cZ \over \partial z} \ .
\label{RfromZ}
\ee
%\alpha==z

\subsection{High density expansion}
\label{HIGHDEN}
Let us first show how this field theory can be used to derive a high density expansion.  

At this end it is convenient to rescale $z$ as $z=\rho \hat z$ and to study the limit $\rho \to \infty$ at
fixed $\hat z$.  Neglecting addictive constants, the action, $S$, can be expanded as:
\be
 {1 \over 2} \sum_a
\int dx dy \ \psih^a(x) f^{-1}(x-y) \psih^a(y)
-{1 \over 2\hat z} \int dx {\sum_a \psih_a(x)^2 } +O(1/(\rho \hat z^{2}) \ ,
\ee

Gathering the various contributions and taking care of the addictive constants, one gets for the resolvent:
\be
\rho R(z)={1 \over z} +    
 {1 \over \rho} \int dk\((
{1\over ( \hat z -\tf(k))} -{1\over \hat z}
\)) \ .
\label{Rleading}
\ee

One can study with this method the eigenvalue density for eigenvalues $|\lam| \sim O(\rho)$, by setting
$\hat z=\lam + i \eps$ and computing the imaginary part of the resolvent in the small $\eps$ limit.  For $\rho \to
\infty$ the leading term  gives a trivial result for the eigenvalue density $ g(\lam)$ i.e.
\be
g(\lam)=\delta(\lam) \ .
\ee
Including the leading large $\rho$ correction that we have just computed, we find that $g(\lam)$ develops,
away from the peak at $\lam \sim 0$, a component of the form:
\be
g(\lam) \sim {1 \over \rho} \int dk \delta ( \lam-\tf(k)) \ .
\label{spec_highrho}
\ee

The continuos part of the spectrum can also be derived from the following simple argument.
We suppose that the eigenvalue $\omega_{i}$ is a smooth function of $x_{i}$. We can thus write:
\be
\sum_{k} f(x-x_{k})\omega(x_{k}) \approx \rho \int dy f(x-y) \omega(y)=\rho \lambda \omega(x) \label{eveq}
\ee
and the eigenvalues are the same of the integral operator with kernel $f$.

This argument holds if the discrete sum in eq.  (\ref{eveq}) samples correctly the continuous
integral.  This will be the case only when the density $\rho$ is large enough that the function
$\omega(x)$ doesn't oscillate too much from one point $x_j$ to a neighboring one.  This condition in
momentum space imposes that the spatial frequency $|k|$ be small enough: $|k| \ll \rho^{1/d}$ (
$\rho^{1/d}$ is the inverse of the average interparticle distance).

The same condition 
is present in the field theory derivation.  We assume that $\tf(k)$ decreases at large $k$, and we call 
$k_M$ the typical range of $k$ below which $\tf(k)$ can be neglected.  Let us consider the corrections of 
order $\rho^{-1}$ in eq.  (\ref{Rleading}).  It is clear that, provided $\hat z$ is away from $0$, the 
ratio of the correction term to the leading one is of order $k_{M}^{d}\rho^{-1}$, and the condition that 
the correction be small is just identical to the previous one.  The large density corrections near to the 
peak $z=0$ cannot be studied with this method.

In conclusion the large $\rho$ expansion gives reasonable results at non-zero $z$ but it does not 
produce a well behaved spectrum around $z=0$. In the next sections we shall see how to do it. 

The reader may wonder if there is a low density expansion: the answer is positive \cite{CGP1}, however
it is more complex than the high density expansion and it cannot be discussed in these notes for lack of
space.

 \subsection{A direct approach}
Let us compute directly what happens in the region where $\rho$ is very large.
We first make a very general observation: if the points $x$ are random and uncorrelated we have that
\bea
\sum_{i}A(x_{i})= \rho \int dx A(x)\ ,\\
\sum_{i,k}B(x_{i},x_{k})= 
\sum_{i,k;i\ne k}B(x_{i},x_{k})+\sum_{i}B(x_{i},x_{i})=\nn \\
\rho^{2} \int dx dy B(x,y)+ \rho \int dxB(x,x)
\eea
When the density is very high the second term can be neglected  and we can 
approximate multiple sum with multiple integrals, neglecting the contribution coming from coinciding 
points \cite{MPZ,MPg,LONG}.

We can apply these ideas to the high density expansion of the Green function in momentum space. Using a simple
Taylor expansion in $1/z$ we can write:
\be
 G(p,z)=\frac1{z}\sum_{R}(-z)^{-R }\overline{M^{R}(p)}
\ee
where for example
\be
M^{3}(p)=N^{-1}\sum_{k_{0},k_{1}k_{2},k_{3}}f(x_{k_{0}}-x_{k_{1}})f(x_{k_{1}}-x_{k_{2}})
f(x_{k_{2}}-x_{k_{3}})\exp(ip(x_{k_{0}}-x_{k_{3}})
\ee
The contribution of all different points gives the large $\rho$ limit, and taking care of the contribution of
pairs of points that are not equal gives the subleading corrections.  In principle everything can be computed
with this kind of approach that is more explicit than the field theory, although the combinatorics may become
rather difficult for complex diagrams and the number of contributions become quite large when one consider
higher order corrections \cite{V1,LONG}.  

Generally speaking the leading contribution involve only convolutions and is trivial in momentum space, while
 at the order $1/rho^m$, there are $m$ loops for diagrams that have $m$ intersections: the high $\rho$
expansion is also a loop expansion.

 \subsection{A variational approximation}
In order to  elaborate a general approximation for the spectrum, that should produce sensible results for 
the whole spectrum, we can  use a standard Gaussian
variational approximation in the field theory representation, that, depending on the context, appears
under various names in the literature, like Hartree-Fock approximation
the Random Phase Approximation or CPA.  

Here we show the implementation in this particular case. By changing $\psih \to i \psih$ in the representation
for the partition function, we obtain:
\be
\cZ= \int D[\psih^a] \exp(S)) \ ,
\ee
where
\bea
S = -{1 \over 2}
\int dx dy
\sum_a \psih^a(x) f^{-1}(x,y) \psih^a(y) \\
+\rho z^{-n/2} \int dx
\exp \(( {1 \over 2 z} \sum_a \psih^a (x)^2 \))  \ .
\eea
We look for the best
quadratic action
\be
S_v=-(1/2) \sum_{ab} \int dx dy G^{-1}_{ab}(x,y) \psih^a(x) \psih^b(y)
\ee
that approximates the full interacting problem.

This procedure is well known in the literature and it gives the following result.  If we have a field theory
with only one field and 
with action
\be
\int dp D(p) \phi(p)^{2} +\int dx V(\phi(x)),
\ee
the propagator is given by
\be
G(p)={1\over D(p)+\Sigma}\ ,
\ee
where 
\be
\Sigma= \lan V''(\phi) \ran_{G} \ .
\ee
In other words $\Sigma$ is a momentum independent self-energy that is equal to the expectation value of
$V''(\phi)$ in theory where the field $\phi$ is Gaussian and has a propagator equal to $G$.

In the present case, there are some extra complications due to the presence of indices; the appropriate form of
the propagator is $G_{ab}(p)=\delta_{ab} G(p)$.  After same easy computations, one finds that $\tilde G(p)$
satisfies the self consistency equation:
\be
 \tilde G (p) = { 1\over f^{-1}(p)-\Sigma } \ \ , \ \
 \Sigma=-{\rho \over z-\int dk \tilde G(k)}
\label{Ma}
\ee
and the resolvent is given by:
\be
R(z)
={1 \over z-\int dk \tilde G(k)} \ .
\label{Mc}
\ee

Formulas (\ref{Ma},\ref{Mc}) provide a closed set of equations that allow us to compute the Gaussian
variational approximation to the spectrum for any values of $z$ and the density; if the integral would be
dominated by a single momentum, we would recover the usual Dyson semi-circle distribution for fully random
matrices .  In sect.  \ref{num} we shall compare this approximation to some numerical estimates of the
spectrum for various functions $f$ and densities.  The variational approximation correspond to the computation
of the sum of the so called tadpole diagrams, that can be often done a compact way.

The main advantage of the variational approximation is that the singularity of the spectrum at $z=0$ is 
removed and the spectrum starts to have  a more reasonable shape.

Another partial resummation of the $\rho$ expansion can also be done in the following way: if one neglects the
triangular-like correlations between the distances of the points (an approximation that becomes correct in
large dimensions, provided the function $f$ is rescaled properly with dimension), the problem maps onto that
of a diluted random matrix with independent elements.  This problem can be studied explicitly using the
methods of \cite{Abou,diluite}.  It leads to integral equations that can be solved numerically.  The equations
one gets correspond to the first order in the virial expansion, where one introduces as variational parameter
the local probability distribution of the field $\phi$ \cite{CGP}.  The discussion of this point is
interesting and it is connected to the low-density expansion: however it cannot be done here.

An other very interesting point that we will ignore is the computation of the tails in the probability
distribution of the spectrum representation leading to compact expressions for the behavior of the tail
\cite{AZ}.

\subsection{Some numerical experiments}

\label{num}

In this section we present the result of some numerical experiments and the comparison with the theoretical
results (we also slightly change the notation ,i.e. we set $z=\lambda+i \eps$. 

We first remark that for a function $f(x)$ that has a positive Fourier transform
$\tf(k)$, the spectrum is concentrated an the positive axis.
Indeed, if  we call $\omega_i$ a normalized eigenvector  of
the matrix $M$ defined in (\ref{pb}), with eigenvalue $\lambda$, one
has:
\be
\sum_{ij} \omega_i f(x_i-x_j) \omega_j = \lambda \ ,
\ee
and the positivity of the Fourier transform of $f$ implies that $\lambda \ge 0$.

\begin{figure}
\centerline{\hbox{
\epsfig{figure=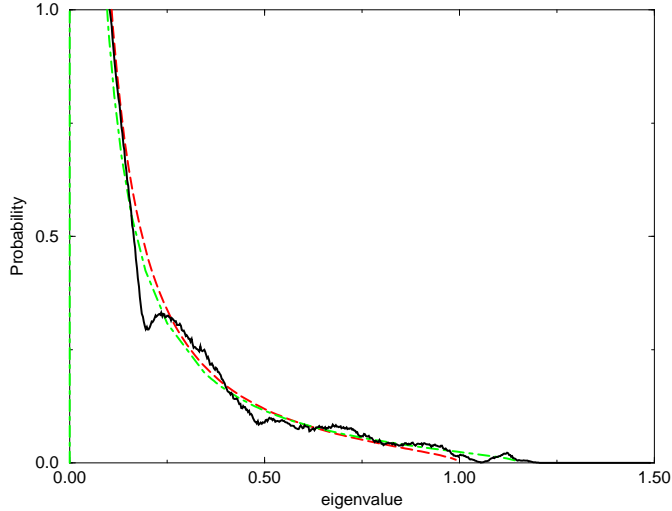,width=8 cm,angle=-90}
}}
\caption{ Density of eigenvalues of a Euclidean Random Matrix in three dimensions,
density $\rho=1$.
The function $f$ is
$f(x)=(2\pi)^{-3/2}\exp(-x^2/2)$, and the matrix is defined from
eq.(\ref{pb}). The full line is the result of a numerical simulation
with $N=800$ points, averaged over $100$ samples. The dashed line
is the result from the high density expansion. The dash-dotted  line is the result 
from the Gaussian variational approximation (RPA) to the field theory.}
\label{fig1}
\end{figure}

We have studied numerically the problem in dimension $d=3$
 with the  Gaussian
function $f(x)=(2\pi)^{-3/2}\exp(-x^2/2)$. 
 In this Gaussian case
the high density approximation gives a spectrum 
\be
g(\lambda) \sim {1 \over \rho \pi^2 } {1 \over \lambda}
\(({1 \over 2} \log{\rho \over \lambda}\))^{1/2} \theta(\rho-\lambda) + C\delta(\lambda)
\label{gauss_highrho}
\ee
Notice that this spectrum is supposed to hold away from the small $\lambda$ peak,
and in fact it is not normalizable at small $\lambda$.

It is possible to do a variational approximation computation  taking care of all the non-linear terms in 
the action for the fields.  This corresponds (as usually) a a resummation of a selected class of 
diagrams. After some computations, one finds that 
one needs to solve,  given $z=\lambda-i\eps$, the following
equations for $C(z)\equiv a(z)+ib(z)$:

\bea
\lambda&=&\rho {a \over a^2+b^2} + {1 \over 2 \pi^2} \int_0^\infty k^2 dk
{e^{k^2/2}-u \over \((e^{k^2/2}-a\))^2 + b^2} \\
\eps &=& \rho {b \over a^2+b^2} -{b \over 2 \pi^2} \int_0^\infty k^2 dk
{1 \over \((e^{k^2/2}-a\))^2 + b^2} \ .
\label{eqgauss}
\eea
One needs to find a solution in the limit where $\eps \to 0$.

\begin{figure}
\centerline{\hbox{
\epsfig{figure=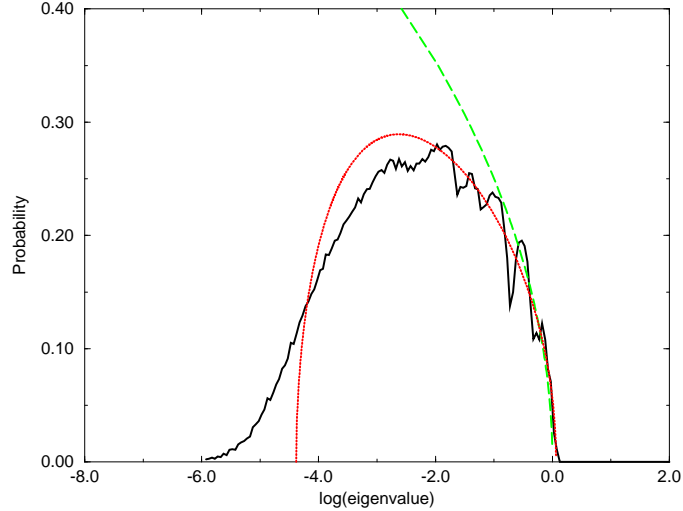,width=8 cm,angle=-90}
}}
\caption{ Density of the logarithm (in base 10) of the
eigenvalues of a Euclidean Random Matrix in three dimensions, density $\rho=1$ (same data of the previous 
figure).
The dashed  line
is the result from the high density expansion. The dotted  line is the result 
from the Gaussian variational approximation (RPA) to the field theory.}
\label{fig2}
\end{figure}

In fig.  (\ref{fig1},\ref{fig2}), we plot the obtained spectrum, averaged over 100 realizations, for $N=800$
points at density $\rho=1$ (We checked that with a different number of points the spectrum is similar).  Also
shown are the high density approximation (\ref{gauss_highrho}), and the result from the variational
approximation.  We see from fig.(\ref{fig1}) that the part of the spectrum $\lambda \in [0.2,1.5]$ is rather
well reproduced from both approximations, although the variational method does a better job at matching the
upper edge.  On the other hand the probability distribution of the logarithm of the eigenvalues
(fig.\ref{fig2}) makes it clear that the high density approximation is not valid at small eigenvalues, while
the variational approximation gives a sensible result.  One drawback of the variational approximation, though,
is that it always produces sharp bands with a square root singularity, in contrast to the tails that are seen
numerically.

In fig.\ref{fig3}, we plot the obtained spectrum,
averaged over 200 realizations, for $N=800$ points at  density $\rho=0.1$.
 We show also a low density approximation (that we do not describe here \cite{MPZ}), and the result from
the variational approximation. We see from fig.(\ref{fig3}) that this
value of $\rho=0.1$ is more in the low density regime, and in particular there
exists a  peak around $\lambda=f(0)$ due to the isolated 
clusters containing small number of points.
The
variational approximation gives the main orders of magnitude
of the distribution, but it is not able to reproduce the details of the
spectrum, in particular the peak due to small clusters.
On the other hand the leading term of a low density
approximation (introduced in \cite{MPZ}) gives a poor approximation the the overall form of the spectrum. One 
should use an approach where the advantages of both methods are combined together.

\begin{figure}
\centerline{\hbox{
\epsfig{figure=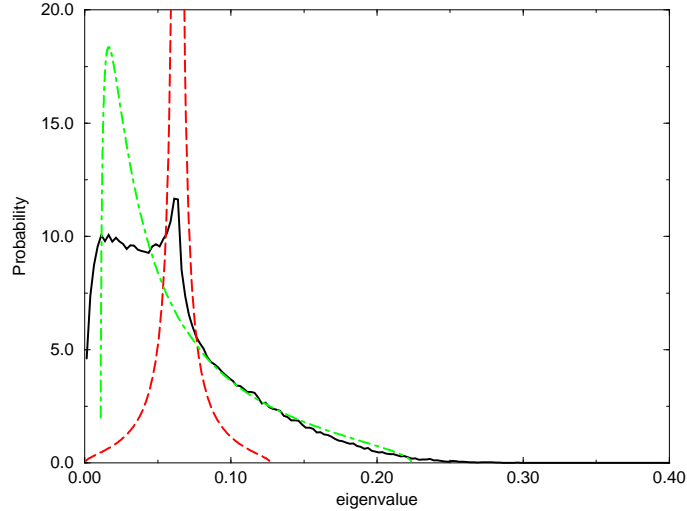,width=8 cm,angle=-90}
}}
\caption{ Density of eigenvalues of a Euclidean Random Matrix in three dimensions,
density $\rho=0.1$.
The function $f$ is
$f(x)=(2\pi)^{-3/2}\exp(-x^2/2)$, and the matrix is defined from
eq.(\ref{pb}) with $u=0$. The full line is the result of a numerical simulation
with $N=800$ points, averaged over $200$ samples. The dashed  line
is the result from a low density expansion introduced in \cite{MPZ}. The dash-dotted line is the result 
from the Gaussian variational approximation (RPA) to the field theory.}
\label{fig3}
\end{figure}

\section{Phonons}
\subsection{Physical Motivations}
Inelastic X-ray scattering (IXS) experiments and inelastic neutron scattering on structural glasses and
supercooled liquids provided useful information on the dynamics of their amorphous structure, at frequencies
larger than $0.1$ THz (see for example \cite{BOSEREV} and references therein).  Those experiments show a
regime, when the wavelength of the plane wave is comparable with the inter-particle distance, where the
vibrational spectrum can be understood in terms of propagation of quasi-elastic sound waves, the so-called
high frequency sound.  This high-frequency sound has also been observed in molecular dynamical simulations of
strong and fragile liquids, and it displays several rather universal features.  In particular, a peak is
observed in the dynamical structure factor at a frequency that depends linearly on the exchanged momentum $p$,
in the region $0.1p_0-1.0p_0$, $p_0$ being the position of the first maximum in the static structure factor.
When extrapolated to zero momentum, this linear dispersion relation yields the macroscopic speed of sound.
The width of the spectral line, $\Gamma$ is well fitted by
\begin{equation}
\Gamma(p)= A p^x\quad ,\quad x\approx 2\,,
\label{BROADENING}
\end{equation}
with $A$ displaying a very mild (if any) temperature dependence.  Moreover, the same scaling of $\Gamma$ has
been found in harmonic Lenhard-Jones glasses~\cite{Ruo2}, and one can safely conclude that the $p^2$
broadening of the high-frequency sound is due to small oscillations in the harmonic approximation.  
In these context other interesting
problems related with the high-frequency vibrational excitations of these topologically disordered
systems~\cite{Ell} regard the origin of the {\em Boson peak} or the importance of localization properties to
understand the dynamics of supercooled liquids~\cite{BeLa}.

The variety of materials where the $p^2$ broadening appears suggests a straightforward physical motivation.
However, the simplest conceivable approximation, a wave propagating on an elastic medium in the presence of
random scatterers, yields Rayleigh dispersion: $\Gamma\propto p^4$.  This result is very robust: as soon as
one assumes the presence of an underlying medium where the sound waves would propagate undisturbed, as in the
disordered-solid model~\cite{HoKoBi,ScDiGa,Montagna}, the $p^4$ scaling appears even if one studies the
interaction with the scatterers non-perturbatively~\cite{MaPaVe}.  When the distinction between the
propagating medium and the scatterers is meaningless (as it happens for topologically disordered systems), the
$p^2$ scaling is recovered.

We want to investigate the problem from the point of view of statistical mechanics of random
matrices, by assuming that vibrations are the only motions allowed in the system.  The formalism
we shall introduce, however, is not limited to the investigation of the high frequency sound and it could be
straightforwardly applied in different physical contexts.

Let us look more carefully at  the relation between vibrational
dynamics in glasses and random matrices. The dynamical structure factor
for a system of $N$ identical particles is defined as:
\begin{equation}
S(p,\omega) = \frac{1}{N} \sum_{i,j} \int dt\ 
 {\mathrm e}^{{\mathrm i}\omega t}\ \left\langle\, {\mathrm e}^{{\mathrm i}  p
\cdot\left( r_j(t)- r_i(0) \right)}\, \right\rangle\, ,
\label{FOURIER}
\end{equation}
where $\langle \dots \rangle$ denotes  the average over the particles
positions $r_j$ with the canonical ensemble.

Here it is convenient to consider the normal modes of the glass
or the supercooled liquid, usually called instantaneous normal modes (INM) because they are supposed to
describe the short time dynamics of the particles in the liquid phase \cite{KEYES}.  One studies
the displacements $u$ around the random positions $x$, by writing the position of the $i$-th particle as
${r}_i(t)= {x}_i + {u}_i(t)$, and linearizing the equations of motion.  Then one is naturally lead to consider
the spectrum of the Hessian matrix of the potential, evaluated on the instantaneous position of the particles.
Calling $\omega_n^2$ the eigenvalues of the Hessian matrix and $ e_n(i)$ the corresponding eigenvectors, the
one excitation approximation to the $S(p,\omega)$ at non zero frequency is given in the classical limit by:
\begin{eqnarray}
S^{(1)}(p,\omega)&=&\frac{k_{\mathrm B}T}{m \omega^2}\,\sum_{n=1}^N\, 
Q_n(p)\,\delta(\omega-\omega_n)\,,\\
Q_n(p)&=&
|\sum_i\, 
 p\cdot e_n(i) \exp({\mathrm i} p\cdot x_i) |^2\,.
\end{eqnarray}

However one cannot always assume that all the normal modes have positive eigenvalues, negative eigenvalues
representing the situation where some particles are moving away from the position $x$.  Indeed, it has been
suggested~\cite{BeLa} that diffusion properties in supercooled liquids can be studied considering the
localization properties of the normal modes of negative eigenvalues.

A related and better defined problem is the study of normal modes at zero temperature, where the displacements
$u$ are taken around the rest positions.  By assuming that this structure corresponds to one minimum of the
potential energy, one can introduce a harmonic approximation where only the vibrations around these minima are
considered, and all the dynamical information is encoded in the spectral properties of the Hessian matrix on
the rest positions.  The Hessian is in a first approximation a random matrix if these rest positions
correspond to a glass phase.  It has been shown using molecular dynamics simulation that below the
experimental glass transition temperature the thermodynamical properties of typical strong glasses
are in a good agreement with such an assumption.

Therefore, the problem of the high-frequency dynamics of the system
can be reduced, in its simplest version,
to the consideration of 
random Euclidean matrices, where the
entries are deterministic functions (the derivatives of the
potential) of the random positions of the
particles. As far as the system has momentum
conservation in our case, due to translational invariance, all the
realizations of the random matrix have a three common normal mode with zero
eigenvalue: the uniform translation of the system.  

An Euclidean matrix
is determined by the particular deterministic function that we are
considering, and by the probabilistic distribution function of the
particles, that in the INM case is given by the Boltzmann
factor. However, for the sake of simplicity we shall concentrate here
on the simplest kind of euclidean matrices without spatial correlations and we will neglect the vector indices
of the displacement. We consider
$N$ particles placed at positions $x_i, i=1,2,...,N$ inside a box, where
periodic boundary conditions are applied. Then, the scalar euclidean matrices
are given by eq.(\ref{TT}), where $f( x)$ is a scalar function depending on the distance between pairs
of particles, and the positions $\{ {x} \}$ of the particles are
drawn with a flat probability distribution.  Notice that the matrix
(\ref{TT}) preserves translation invariance, since the uniform
vector $e_0(i)=const$ is an eigenvector of $M$ with  zero eigenvalue. 
Since there
are not internal indices (the particle displacements are restricted to
be all collinear), we cannot separate longitudinal and transversal
excitations.  

The dynamical structure factor for a scalar Euclidean matrix is given
by
\begin{eqnarray}
S_E(p,E)&=& \overline{\sum_n Q_n(p)\delta(E - E_n)}\,,\label{DEFINIZIONE}\\
Q_n(p)&=& \frac{1}{N}\left|\sum_{i=1}^N\, e_n(i) 
\exp({\mathrm i}  p\cdot x_i)\right|^2\,,\\
S(p,\omega)&=&2\omega S_E(p,\omega^2)\,,
\label{NORMA}
\end{eqnarray}
where the overline stands for the average over the particles position and 
we have given the definition either in the eigenvalue space ($S_E(p,E)$)
and in the frequency space ($S(p,\omega)$). 

\subsection{A more complex field theory representation}

The basic field theory representation is similar to the one of the previous section. The main 
complication is due to the presence of a diagonal term in the matrix $M$. One way out is to introduce one 
more pair of Bosonic fields.
To perform the spatial integrations it turns out to be convenient to represent
the Bosonic fields $\phi$  using new Bosonic fields , i.e.: 
\bq
\chi(x) &\equiv& \sum_{i,a} (\pia)^2 \delta(x-x_i) \\
\psi^{(a)}(x) &\equiv& \sum_i \pia \delta(x-x_i) 
\ ,
\label{CAMPI}
\eq
and using the ``Lagrange multipliers'' $\,\hat{\psia}(x),
\hat{\chi}(x)$, to enforce  the three constraints~(\ref{CAMPI}). 
At this point, the Gaussian variables $\pia$ are
decoupled and can be integrated out.

Skipping intermediate steps and using the fact that all the replicas are
equivalent one can write the Green function as a correlation function:
\bea
G(p,z) = \\
\lim_{n \to 0} \frac{1}{N} \douintx \:
\: e^{i p(x-y)} \int{D [\psi^{(a)}, \hat{\psi^{(a)}}, \chi, \hat{\chi}]
  \: \psi^{(1)}(x) \psi^{(1)}(y)  A^N
\: \exp({S_0}) }\ , 
\label{GIDIPREP}
\eea
where we have introduced the following quantities:
\bea
S_0 \equiv i \sum_{a=1,n} \intx \: \tpsia(x) \psia(x)
+  \intx \chi(x) \hat{\chi}(x) \\
- \frac{1}{2} \sum_{a=1,n} \douintx
\psia(x) f(x-y) \psia(y)
\nn \\&A \equiv \int{ {d^d x}
%\left(\frac{2 \pi}{z+\tchi(x)} \right)^{n/2}
\exp \Phi(x) } \nn \\
\Phi(x) \equiv + i\inty f(y-x) \chi(y) -\frac{1}{2} \frac{\psumma}{z+\hat{\chi}(x)} \ .
\label{CANONICO}
\eea

In order to take easily the limits $N,V \to \infty$ in (\ref{GIDIPREP}) we shall resort to a grand
canonical formulation of the disorder, introducing the partition function ${\cal Z}[\rho] \equiv
\sum_N A^N \rho^N/N! \: \exp{S_0}$.  Since the average number of particles $\overline{N} = V \rho
\langle A \rangle$, in the $n \to 0$ limit we have that $\langle A \rangle=1$ and the 'activity'
$\rho$ is just the density of points, as before.  Furthermore, the Gaussian integration over the
fields $\psia$ is easily performed, leading to the field theory:
\be
{\cal Z}\left[\rho, z\right] = \int{D [\hat{\psi^{(a)}}, \chi, \hat{\chi}]
\exp{(S'_0 + S_1)} } \ ,
\label{GRANCAN}
\ee
where \footnote{The quantity $\ln f$ is computing considering $f$ as an integral operator}
\bea
S'_0 \equiv -\frac{n}{2} \tr \ln f + \intx \chi(x) \hat{\chi}(x) -  //
\frac{1}{2} 
\sum_{a=1,n} \douintx \: \tpsia(x) f^{(-1)}(x-y) \tpsia(y)  
\label{INTERAGENTE} \\
S_1 \equiv \rho \intx \: \left( \exp{\Phi(x)}  \right)\nn
\eea
The resolvent  is related to  the correlation function  by eq.(\ref{GFINALE}).
Before  computing (\ref{GFINALE}), let us turn to the
symmetry due to the translational invariance. Since $e_n(i)=constant$
is an eigenvector of zero eigenvalue of the matrix~(\ref{TT})
for every disorder realization, we see that Eq.(\ref{GDIPI}) implies:
\be G(p=0,z) = \frac{1}{z} \  .
\label{IMPULSONULLO}
\ee
Interestingly enough, in the framework of the field theory introduced
above, that constraint is automatically satisfied, due
to the Ward identity linked to that symmetry.
The interaction term $\Phi(x)$ is indeed  invariant under the
following infinitesimal transformation of order $\eta$:
\bq
\delta \tpsi(x) = i \eta \left( z+\tchi(x) \right)\, , \\
\delta \chi(x) = 2 i \eta \inty f^{-1}(x-y) \tpsi(y) 
\left( z+\tchi(x) \right) \, , 
\label{TRASFO}
\eq
while the whole variation of the action
$S=S'_0+S_1$ is due to the non-interacting part
$S'_0$:
\be
\delta S = -i\eta \frac{z}{\f0} \intx \tpsi(x)
\label{VARIAZIONE}
\ee
The invariance of $S_1$ hence leads to the following Ward identity:
\be
\langle z + \tchi(x) \rangle = \frac{z}{\f0} \langle
\tpsi(x) \inty \: \tpsi(y) \rangle
\label{WARD}
\ee
That is enough to prove that 
\be
\inty \tpsi(0) \tpsi(y)
\ee
diverges as $1/z$ at small zeta.
If we combine the previous result  with the exact relation
$\langle \tchi(x) \rangle = -\rf0$, that can be derived  with a different argument, to obtain:
\be
(z - \rf0)\frac{\f0}{z} = \inty \: \langle \tpsi(x) \tpsi(y) \rangle \ `
\label{WARD2}
\ee
that, together with (\ref{GFINALE}), implies the expected constraint
$G(p=0,z)=1/z$.

The interacting term in (\ref{INTERAGENTE}) is complicated by the
presence of an exponential interaction, meaning an infinite number of
vertices. In order to perform the explicit computation of the
resolvent $G(p,z)$  one has to introduce some scheme of
approximation.  We have chosen to deal with the high density limit,
where many particles lie inside the range of the interaction $f(r)$.
The high density limit ($\rho \gg 1$) of (\ref{GFINALE}) is the
typical situation one finds in many interesting physical situations,
for example the glassy phase.  In order to extract the leading term
let us make the expansion $\exp \Phi(x) -1 \sim \Phi(x)$.  In that
case the integration over the fields $\chi, \tchi$ is trivial, because
of: 
\bea
\int D [\chi] \exp \left( \intx \chi(x) \left[ \rho \inty f(x-y) +
\tchi(x) \right] \right)  = \\
\prod_x \delta(\tchi(x) + \rf0)
\label{DELTA}
\eea
and the fields $\tpsia$ are free.
In fact, introducing the quantity $a \equiv z - \rf0$, one remains with the
Gaussian integration:
\be
{\cal Z} \propto \int D[ \tpsia] \exp{-\frac{1}{2} \douintx \sum_a \tpsia(x) K^{-1}(x-y)
\tpsia(y)} \ ,
\label{SEMPLICE}
\ee
where the free propagator $K$ is defined by:
\be
K^{-1}(x-y) \equiv f^{-1}(x-y) + \frac{\rho}{a} \delta(x-y) \ .
\label{KAPPA}
\ee
It is then easy from (\ref{GFINALE}) and (\ref{SEMPLICE}) to obtain
the result:
\begin{equation}
G_0(p,z) = \frac{1}{z-\ep}\, ,
\end{equation}
where 
\be
\ep=\sqrt{\rho(\tilde f(p)- \tilde f(0)}
\ee
We see that at the leading
order, a plane wave with momentum $p$ is actually an eigenstate of the
matrix $M$ with eigenvalue $E$, and the disorder does not play any
relevant role. In other words, inside a wavelength $2 \pi /p$ there is
always an infinite number of particles, ruling out the density
fluctuations of the particles: 
the system reacts as an elastic medium.

Let us finally obtain the density of states at this level of accuracy, using
Eq(\ref{DOF}) and $G_0$:
\begin{equation}
g_E(E)=\delta\left(E-\rho\ff(0)\right)\,.\label{EINSTEIN}
\end{equation}
We obtain a single delta function at $\rho\ff(0)$, that is somehow
contradictory with our result for the dynamical structure factor: from
the density of states one would say that the dispersion relation is
Einstein's like, without any momentum dependence! The way out of this
contradiction is of course that in the limit of infinite $\rho$ both
$\ep$ and $\rho\ff(0)$ diverge. 
The delta function in eq. (\ref{EINSTEIN}) is the leading term in $\rho$, 
while the states 
that contribute to the dynamical  structure factor
appear only in the subleading terms in the 
density of states. The same phenomenon is present in the simpler case discussed in the previous section.

subsection{One loop}\label{ONELOOPSECT}

We have seen above that, with the expansion of
$\exp \Phi(x)$ up to first order in $\Phi(x)$ the fields $\hat{\psi}$ are
non interacting and no self energy is present. Now we shall see that
the one-loop correction to that leading term provides the $\frac{1}{\rho}$
contribution to the self energy. In fact by adding the quadratic term
to $S_1$ the total action becomes (in the $n \to 0$ limit):

\bq
S = S'_0+S_1 \sim S'_0 + \rho \intx \: \left( \Phi(x) +\frac{1}{2}
\Phi^2(x) \right) \nn \\
= -\frac{\rho}{2} \douintx \: \chi(x) f^{(2)}(x-y) \chi(y) +
i \intx \: b(x) \chi(x) \nn \\
- -\frac{1}{2} \sum_{a=1,n} \douintx \: \tpsia(x) c(x-y) \tpsia(y) \nn \\
+ \frac{\rho}{8} \intx \: \left( \frac{\psumma}{z+\tchi(x)} \right)^2 \ ,
\label{PHI2}
\eq
where
\bq
b(x) &\equiv& \tchi(x) + \rf0 - \frac{\rho}{2} \inty \: \frac{\psumma}
{z+\tchi(x)} f(y-x) \nn \\
c(x-y) &\equiv& f^{(-1)}(x-y) + \frac{\rho}{z+\tchi(x)} \delta(x-y)
\label{DEFINIZIONI}
\eq
After doing some computations and adding the two contributions coming two different diagrams one gets
\bea
G(p,z) =\\
G_0(p,z) + G_0^2(p,z) \frac{1}{\rho}
\unoint \: G_0(q,z) \left( \diffe{q} \right)^2
\label{ONELOOP}
\eea
The Dyson resummation of all the higher orders terms, that is
built by 'decorating' recursively all the external legs $G_0$ with the
one loop correction in (\ref{ONELOOP}) gives:
\be
G(p,z) = \frac{1}{z-\eps(p)- \Sigma_1(p,z)} \ ,
\label{DYSON1}
\ee
where the self-energy $\Sigma_1(p,z)$ is given by
\be
\Sigma_1(p,z) \equiv \frac{1}{\rho}\unoint \: G_0(q,z) \diffe{q}^2 \ .
\label{SELF1}
\ee
Let us study in details the low exchanged momentum limit of 
Eq.(\ref{SELF1}). It is clear that at $p=0$ the self-energy vanishes,
as required by the Ward identity (\ref{WARD}). 
We need to expand $\ff(p-q)$ for small $p$, that due to the spherical
symmetry of $\ff$ yields
\begin{eqnarray}
\ff(p-q)&=&\ff(q)- ( p\cdot  q)\, \frac{\ff'(q)}{q}+{\cal O}(p^2)\,,\\
&=&\ff(q)+( p\cdot q)\,  \frac{\epsilon'(q)}{q\rho}+{\cal O}(p^2)\,.
\label{PICCOLOP}
\end{eqnarray}

Substituting (\ref{PICCOLOP}) in (\ref{SELF1}), and performing
explicitly  the trivial angular integrations in dimensions $d$ we obtain
\bea
\Sigma_1(p,z)\approx p^2 \frac{2^{1-d}}{\rho d \pi^{d/2}\Gamma(d/2)}
\int_0^\infty dq\,q^{d-1}\frac{\left[\epsilon'(q)\right]^2}{z-\epsilon(q)}
=
\\
p^2 \frac{2^{1-d}}{\rho d \pi^{d/2}\Gamma(d/2)}
\int_0^{\epsilon(q=\infty)} d\epsilon\,
\frac{\left[q(\epsilon)\right]^{d-1}}{q'(\epsilon)(z-\epsilon)}\,.
\eea
In the last equation, we have denoted with $q(\epsilon)$ the inverse
of the function $\epsilon(q)$.  Setting now $z= E+{\mathrm i} 0^+$,
and observing that $\ep\approx A p^2$ for small $p$, we readily obtain
\begin{eqnarray}
Re \Sigma_1(p, E+{\mathrm i} 0^+)&\approx& 
p^2 \frac{2^{1-d}}{\rho d \pi^{d/2}\Gamma(d/2)} 
\int_0^{\epsilon(q=\infty)} d\epsilon\,
\frac{\left[q(\epsilon)\right]^{d-1}}{q'(\epsilon)(E-\epsilon)}\,,
\label{REASSI1L}
\\
Im \Sigma_1(p, E+{\mathrm i} 0^+)&\approx& -
 \frac{\pi 2^{2-d} A}{\rho d \pi^{d/2}\Gamma(d/2)} p^2 [q(E)]^d\,.
\label{IMASSI1L}
\end{eqnarray}
Since the principal part is a number of order one,
the real part of the self-energy  scales like $p^2$ (possibly
with logarithmic corrections), and thus the speed of sound of the
system renormalizes due to the $1/\rho$ corrections. As a consequence,
the function $q(E)$ is proportional to 
$E^{1/2}\sim p$ at the maximum of the function of $p$
$S_E(p,E)$, and the width of the peak of the $S_E(p,E)$ will scale like
$p^{d+2}$. It is then easy to check (see (\ref{NORMA}))
 that in frequency space the width
of the spectral line will scale like 
\be 
\Gamma \propto p^{d+1}\,,
\label{FREQUENZE}
\ee
as one would expect from Rayleigh scattering considerations. 

The result (\ref{FREQUENZE}) for the asymptotic regime $p<<1$ has been found at the one loop level.
In order to predict correctly the spectral properties at very low external momentum $p$, it turns
out that one must study the behavior of the two loop contribution, that can be done in details.
Nevertheless, the one loop result is already a good starting point to perform detailed comparisons
with the numerical simulations.  The disadvantage of this approach is that it works near band edge
($\omega=0$ at high $\rho$ but is not suited for producing the whole spectrum.  Due to the
complication of the action it is not clear how to do a variational computation and the best it can
be done at the present moment is a CPA-like approximation described in the next section.

\subsection{A CPA like approximation}

As in the previous section we consider the resolvent $G(p,z) $
Our aim is to compute $G$ using the appropriate self-consistent
equations.    A partial resummation of the 
 expansion  for the resolvent can be written as
\begin{equation}
G(p,z) =  \frac{1}{z-\lambda(p)-\Sigma(p,z)}\,.\label{GDIPI1}
\end{equation}
The self-energy $\Sigma(p,z)$
is then expanded in powers of $1/\rho$ \cite{MPZ,LONG} in the
relevant region where $\rho=O(z)$. 

\begin{figure}
\epsfig{file=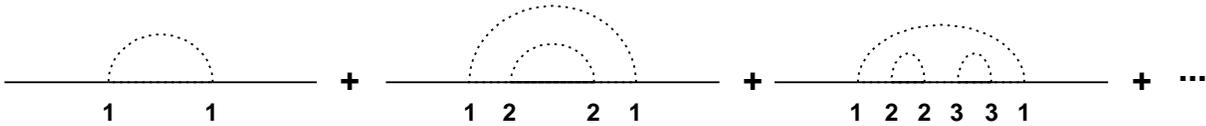, angle=0, width=\columnwidth}
\caption{The diagrams of the $1/\rho$ expansion that are taken into
account in our approach. The numbers correspond to the particle-label
repetitions}
\label{DIAGRAMMI}
\end{figure}

If we reformulate the $1/\rho$ expansion in a diagrammatic way
we can  identify those diagrams
with the simple topology of Fig.~\ref{DIAGRAMMI}. Topologically, these
diagrams are exactly those considered in the usual lattice CPA and in
other self consistent approximations.  The sum of this infinite subset
is given by the solution of the integral equation:
\begin{equation}
\Sigma(p,z)=\frac{1}{\rho}\int \!\! \frac{d^3 q}{(2\pi)^3} 
\left[\rho\left(\hat f(\mbox{\boldmath$q$})-\hat f(\mbox{\boldmath$p$}-
\mbox{\boldmath$q$})\right)\right]^2 G(p,z),
\label{CACTUS}
\end{equation}
where the resolvent is given by Eq.~(\ref{GDIPI1}). The solution gives
us the resolvent, and hence the dynamical structure function and  density of states (Eq.~\ref{PGRANDE} below).

We are interested to study the solution of Eq.~(\ref{CACTUS}) for different
values of $z$ and $\rho$.  To be definite, we consider an explicit
case where the function $f(r)$ has a simple form, namely
$f(r)=\exp[-r^2/(2\sigma^2)]$.  This is a reasonable first
approximation for the effective interaction~\cite{Proceedings}.  We
shall take $\sigma$ as the unit of length and set $p_0=1/\sigma$,
that is a reasonable choice for $p_0$ for this Gaussian $f(r)$, as
discussed in~\cite{Proceedings}.  In this particular case we will
solve numerically the self-consistence equation.  We will also
evaluate by simulation (using the method of moments~\cite{MOMENTI})
the exact dynamical structure function and the density of states by computing the resolvent for concrete
realizations of the dynamical matrix, considering a sufficiently high
number of particles so that finite volume effects can be neglected.
These numerical results will be supplemented by analytic results, that
are $f$-independent and can be obtained in the limits $p\to\infty$ and
$p\to 0$.

The infinite momentum limit is particularly interesting because of the
remarkable result~\cite{V1,LONG} that  the density of states $g(\omega)$ can be
written as
\begin{equation}
	 g(\omega) = \lim_{p\to\infty} \frac{ \omega^2 S^{(1)}(p,\omega) }
{k_{\mathrm B} T p^2}.
\label{PGRANDE}
\end{equation}
\begin{figure}
\epsfig{file=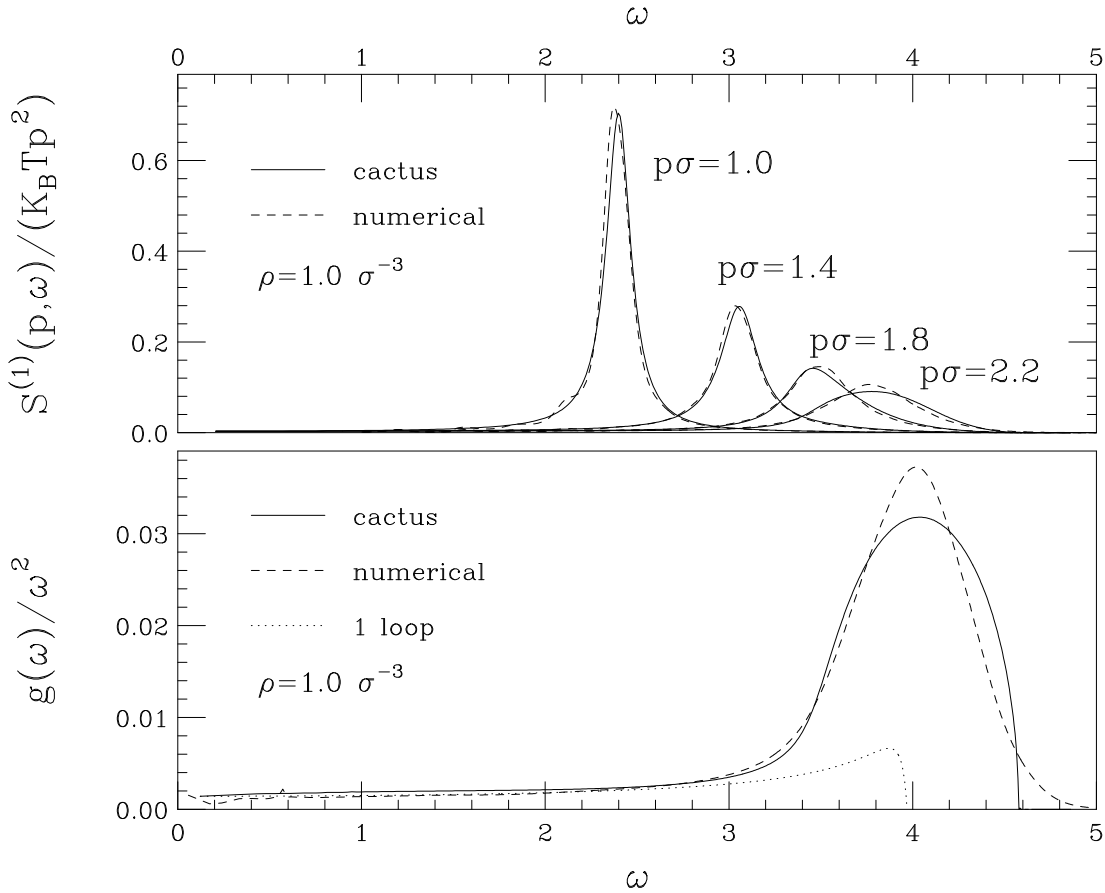, angle=90, width=0.9 \columnwidth}
\caption{Top: dynamic structure factor as obtained from
Eq.(\protect\ref{CACTUS}) (full line) and from
simulations~\protect\cite{LONG} (dashes). Bottom: the density of states divided by
$\omega^2$ (Debye behavior) as obtained from
Eq.(\protect\ref{DENSIDADDEESTADOS}) (full line), simulations
(dashes), and first order in the $1/\rho$ expansion (dots).}
\label{RHO1}
\end{figure}

We easily find that in this limit Eq.~(\ref{CACTUS}) can be written as:
\begin{equation}
\frac{1}{\rho{\cal G}(z)}=\frac{z}{\rho}-\hat f(0) -{\cal A} {\cal G}(z) -\int
\!\!\! \frac{d^3q}{(2\pi)^3}\hat f^2(\mbox{\boldmath$q$})
G(\mbox{\boldmath$q$},z) \ ,
\label{DENSIDADDEESTADOS}
\end{equation} 
where
\be
{\cal G}(z)=\lim_{p\to\infty} G(p,z)
\ee 
and
\be
{\cal A}= (2\pi)^{-3}\int
\!\!  \hat f^2(\mbox{\boldmath$q$})\,d^3q \ .
\ee 

A simple approximation consists in neglecting the last term in the r.h.s. of (\ref{DENSIDADDEESTADOS}),
that is reasonable at large $z$.  This approximation implies a the density of states that is semicircular as a
function of $\omega^2$, with width proportional to $\sqrt{\rho}$ and centered at $\omega^2=\rho\hat f(0)$.
Translational invariance also requires low-frequency modes.  These are given by the neglected term, and in
fact it is easy to show that at high density it produces a Debye spectrum that extends between zero frequency
and the semicircular part.

In the limit $p\to0$, the leading contribution to $\Sigma''$ comes
from $q \gg p$ in Eq.~(\ref{CACTUS}), where $G(q,z) \approx {\cal
G}(z)$, so we can write for the peak width $\Gamma(p) \approx
\Gamma_0(p)$, where
\begin{equation}
\Gamma_0(p) \equiv
\pi \rho \frac{g(\omega_p)}{2\omega_p^2}
\int \!\!\! \frac{d^3 q}{(2\pi)^3}
\left[\hat f(\mbox{\boldmath$q$})-\hat f(\mbox{\boldmath$p$}-
\mbox{\boldmath$q$})\right]^2.
\label{sigma-infinito}
\end{equation}
The integral is of order $p^2$, so if the spectrum is Debye-like for
small frequencies, we get $\Gamma(p) \sim p^2$.

These considerations are verified by the numerical solution of the Gaussian case, that are shown in
Fig.~\ref{RHO1} for $\rho=1.0\, \sigma^{-3}$ together with the results for the simulations~\cite{LONG}.
Note the good agreement, to be expected for high-densities, and how, for large $p$, $S^{(1)}(p,\omega)$
(Fig.~\ref{RHO1}, top) tends to the density of states.  The density of states from the self-consistent
equation (Fig.~\ref{RHO1},bottom) also agrees very well with the results from simulations, and is a big
improvement over the first term of the expansion in powers of $\rho^{-1}$.  The two contributions (Debye and
semicircle) mentioned above can be clearly identified.  As expected our approximation fails in reproducing the
exponential decay of the density of states at high frequencies, that is non perturbative in
$1/\rho$~\cite{Zee2} and corresponds to localized states.

\begin{figure}
\epsfig{file=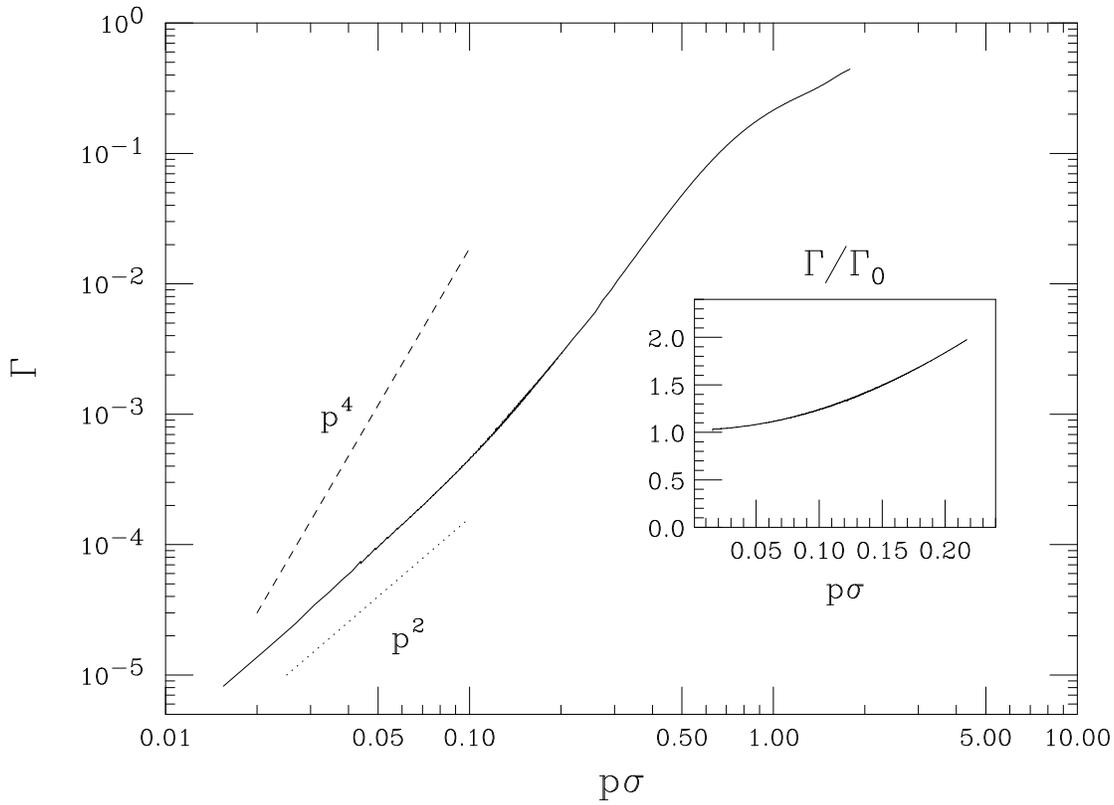, angle=90, width=0.9\columnwidth}
\caption{Peak width vs.\ $p$, for $\rho=0.6\sigma^{-3}$. The inset
shows that $\Gamma_0(p)$ is the dominant contribution at small $p$.}
\label{scaling}
\end{figure}

Next in Fig.~\ref{scaling} we plot the linewidth as a function of $p$ as obtained from Eq.(\ref{CACTUS}).
Notice that we recover the behavior predicted from the first two non-trivial terms in the expansion in powers
of $\rho^{-1}$~\cite{LONG}: the linewidth is proportional to $p^2$ at small $p$ (also predicted by the
argument above), then there is a faster growth and finally it approaches to a constant as $S^{(1)}(q,\omega)$
starts to collapse onto the density of states.  The inset shows that the contribution
Eq.~(\ref{sigma-infinito}) is indeed dominant at small $p$.  However, accurate $p^2$ scaling is found only for
very small momenta ($p/p_0<0.1$), while experiments are done at $0.1<p/p_0<1$.  In this crossover region, our
approach predicts the existence of non-universal, model dependent small deviations from $p^2$, that are
probably hard to measure experimentally.  In any case, the effective exponent is certainly less than 4, in
contrast with lattice models and consistent with experimental findings.  Similar conclusions can be drawn from
mode coupling theory (see fig.~8 of \cite{GoMa}).

\subsection{The disappearance of phonons}

New phenomena are present the function $f$ is no more positive or in the vectorial case.
In these case it is possible that the spectrum arrives up to negative values and the density of states 
$g_{E}$ does not vanish at $E=0$. In principle for a given model we should not expect a sharp transition 
from the situation where the Debye spectrum holds and $g_{E}(E)\propto E^{1/2}$ because tails are always 
present. However often tails are small and one can effectively observe a transition from the two regimes.

In the framework where the density of states is computed from a simple integral equation, the tails are 
neglected and we are in the best situation to observe such  a phenomenon. 

Let us call $\tau$ a parameter that separate the two regimes. A detailed analysis show that a $\tau=0$ the 
density of states behaves as 
\be
g_{E}(E)\propto E^{1/4},
\ee
while at small $\tau$ and $E$ we have
\be
g_{e}(E,\tau)=E^{1/4}h(\tau E^{1/2}).
\ee
The detailed argument is a delicate but let us consider an heuristic version of it.

With some work it is possible to obtain
from~(\ref{CACTUS}) an integral equation even for  the density of states.
As a matter of fact, defining ${\cal G}(z) = G(p=\infty,z)$, the  density of states turns
out to be
\begin{equation}
g_{E}(E) = - {2\over \pi} \, \mathrm{Im}\, {\cal G} (E
+ i 0^+).
\end{equation} 
$\cal G$ being the solution of the following  equation:
\begin{equation}
\frac{1}{\rho{\cal G}(z)}=\frac{z}{\rho}-\hat f(0) - A {\cal G}(z) -B(z)
\label{DENSIDADDEESTADOS1}
\end{equation} 
where
\bea
A= (2\pi)^{-3}\int \!\!\hat f^2(\mbox{\boldmath$q$})\,d^3q \ , \\
B(z)=\int
\!\!\! \frac{d^3q}{(2\pi)^3}\hat f^2(\mbox{\boldmath$q$})
G(\mbox{\boldmath$q$},z) \ ,
\eea
With this equation, one needs to know the resolvent at all $q$ to
obtain the  density of states, due to the last term in the r.h.s. This can be done by
solving numerically the self-consistent equation of
ref.~\cite{Grigera01}, but here we perform an approximate analysis,
that is more illuminating. 

The solution of the previous equation is
\be
{\cal G}(z)={-\alpha(z)+\sqrt{\alpha(z)^{2}-4A\rho}\over 2 A \rho} \ , 
\ee
where 
\be
\alpha(z)= \hat f(0)-\frac{z}{\rho}+B(z)
\ee
The crudest approximation is to neglect the dependence of $B(z)$ from $z$ 
(we will also assume that $B$ is a smooth function of the other parameters).
In this case Eq.~\ref{DENSIDADDEESTADOS1} is quadratic in
${\cal G}$, and one easily finds a semicircular  density of states. 
But
the semicircular spectrum misses the Debye part, and a better
approximation is needed.  So we substitute $G$ in the last term of the
r.h.s. by the resolvent of the continuum elastic medium $G_0(z,p) = (
z - E(p))^{-1}$.  This is reasonable because the $f^2(q)$
factor makes low momenta dominate the integral, and due to
translational invariance $G(z,p) \approx G_0(z,p)$ in this
region~\cite{Grigera01}. We shall be looking at small $E$,
so to a good approximation
\begin{equation}
B(z)=B_{0}+i B_{1}z^{1/2}
\end{equation}
We have two limiting cases.  depending on the sign of  of $\alpha(0)^{2}-4A\rho$.
\begin{itemize}
	 \item
When the
semicircular part of the  density of states does not reach low frequencies, the
square root can be Taylor-expanded, and one gets
\begin{equation}
g_{e}(E) \approx E^{1/2},
\end{equation}
that is precisely Debye's law. 
\item In the opposite situation, on the other hand,
the semicircle arrives also at negative values of $E$ and $g_{E}(E)$ is different from zero also if
$B_{1}$ where zero.
\item Exactly at the critical point we get $g_{E}(E) \propto \sqrt{\sqrt{E}}=E^{1/4}$ that is the announced 
result.
\end{itemize}
Mathematically, the instability arises when ${\cal G}(0)$ develops an 
imaginary part. This can only come from the square root in 
previous equations.
 Notice that this instability is a kind of
phase transition, where the order parameter is $\mathrm{Im}\,
{\cal G}(0)$. Doing a detailed computation on finds that this order parameter
behaves as $\tau^{\beta}$, with $\beta=1/2$.

Since the behavior of the propagator at high momentua does not
strongly affect the dispersion relation~\cite{Grigera01}, we do not
expect deviations from a linear dispersion relation.
This has been checked either numerically or solving numerically the
self-consistency equation given a particular choice for the function
$f(r)$ (see the numerical results section below). It can be argued that this kind of phenomenon is 
responsible of the Boson peak \cite{BOSE}, however we cannot discuss this point for lack of space.

section{Correlated points}
\subsection{Various models}

When the points are correlated things become more difficult.  Already it is difficult to study the
statistical properties of correlated points and it is more difficult to study the properties of the
matrices that depends on these points.  Although some approaches have been developed that allow us
to deal with two points correlations \cite{V1,LONG}, it is not evident how to treat the general case
where many points correlations are present.

A particular interesting case is when the matrix and the distribution of the points are related.
In the best of the possible words there should be extra symmetries that express this relations.

This field is at its infancy, so that I will only describe some general results, without presenting 
applications, that for the moment do not still exist for the case of Euclidean random matrices.

In the general case we consider an Hamiltonian $H[x]$, such that the stationary equations 
\be
F_{i}[x]\equiv{\partial H\over \partial x_{i}}=0
\ee
have a large number of solutions.

Let us label these solutions with a index $\alpha$.  The probability $P[x]$ of a configuration of the points
$x$ is assumed to be given by
\bea
P[x]\propto \sum_{\alpha}w_{\alpha}\delta(x-x^{\alpha})\,,\\
w_{\alpha}\equiv D[M^{\alpha}] )\exp(-\beta H[x^{\alpha}])\,,
\eea
where the matrix $M$ is given by
\be
M^{\alpha}_{i,k}={\partial ^{2}H \over \partial x_{i}x_{k}}|_{x=x^{\alpha}} \ .
\ee
This problem arises for the first time in the framework of spin glasses \cite{MPV}, but its relevance to
glasses has been stressed for the first time in \cite{ParisiAndalo}
The function $F$ may selects the different type of stationary points. 

Different interesting  possibilities are:
\begin{itemize}
	 \item $D[M]=1$, i.e. all stationary points have the same weight.
	 \item $D[M]=\sign({\det[M]})$, i.e. all stationary points have a weight that can be 1 or -1.
	 \item $D[M]=1$ only if all the eigenvalues are positive, otherwise it is zero (i.e. minima are 
	 selected)
\end{itemize}
We are eventually interested to study the properties of the matrix 
\be
1 \over M[x]-z \, ,
\ee
when the points $x$ are extracted with the previous probability.
\subsection{A new supersymmetry}
For simplicity I will only restrict myself to the case $z=0$ where some symmetry are present when 
$D[M]=\sign({\det[M]})$.

Indeed it is evident that
\be
\int P[x]d[x] A[x] \propto
\int d[x] D[M] \exp(-\beta H[x])\det M[x]|\prod_{i}\delta (F_{i}[x]) A[x]
\ee
The last term simplify to 
\be
\int d[x]  \exp(-\beta H[x])\det M[x]\prod_{i}\delta (F_{i}[x]) A[x] \, ,
\ee
when $D[M]=\sign(\det(M))$. For simplicity let us assume that this is the case, without discussing the 
physical motivations of this choice.

Using usual representations we can write
\be
\overline{M^{-1}_{i,k}}=\int d \mu [x,\lambda,\psi, \overline \psi]\; \overline \psi_{i}\psi_{k} \, 
,
\ee
where $d \mu$ is a normalizes measure proportional to 
\be
d[x]d[\lambda]d[\psi]d[\overline \psi] \exp\(( -\beta H[x] +i \sum_{k}\lambda_{k}F_{k}[x] +\sum_{k,j}
M_{j,k}\overline \psi_{j}\psi_{k}\)) \label{action} \ .
\ee
Here the $\psi$ are really Fermionic variables (i.e. anticommuting Grassmann variables): they have
been introduced for representing the determinant, however they can also be used to compute the
matrix elements of the inverse of the matrix $M$.

It was rather unexpected \cite{juanpe,CGPM} to discover that also for $\beta\ne 0$  the measure $d \mu$ 
is invariant under a transformation of Fermionic character of BRST type (a supersymmetry in short).  
(see for example \cite{BRST,zinnjustin}).  If $\epsilon$ is an infinitesimal Grassmann parameter, it 
is straightforward to verify that (\ref{action}) is invariant under the following transformation, 
\beq 
\delta x_i = \epsilon\, \psi_i \quad\quad 
\delta \lambda_i = -\epsilon \, \beta \, \psi_i \quad\quad 
\delta  \bar\psi_i = -\epsilon\, x_i \quad\quad 
\delta \psi_i = 0 \quad\quad \quad 
\label{brst}
\eeq

This symmetry is extremely important and it is crucial to use approximation methods that do not break it
\cite{jorge1}\cite{discuss}. This has been stressed in the spin glass case \cite{CGPM} in the context of the Tap
equations \cite{TAP}.

\subsection{Physical meaning of the supersymmetry} 
\label{ss:BRST}

In this section we will follow a reasoning allowing for an intuitive
explanation of the physical meaning of the supersymmetry
in terms of a particular behavior of the solutions of the stationary equations \cite{CLPR}-\cite{CR}.

It may be interesting to concentrate the attention on some of the Ward identities that are generated by 
the supersymmetry. The simplest one is
\be
\lan \overline \psi_{j} \psi_{k}\ran =i \lan \lambda_{j} x_{k} \ran \label{WARDS} \ .
\ee

Apparently the equation is trivially satisfied. Indeed it is convenient to consider a slightly modified 
theory where we make the substitution
\be
F_{j} \to F_{j}-h \ .
\ee
The final effect is to add an extra term in the exponential equal to $i \lambda_{j} h$. In other words
the r.h.s of equation \ref{WARDS} is  
\be
{\partial \lan x_{k} \ran_{h} \over \partial h}|_{h=0}
\ee

The l.h.s can also be computed and one finds that eq. (\ref{WARDS}) becomes
\be
\sum_{\alpha}w_{\alpha}{\partial \lan x^{\alpha}_{k} \ran_{h} \over \partial h}|_{h=0}=
{\partial \sum_{\alpha'} 
w_{\alpha' }\lan x^{\alpha'}_{k}\ran \over \partial h}|_{h=0}\ .
\ee
If we consider solutions where the $\det (M) \ne 0$ we have that 
\be
{\partial w_{\alpha}\over \partial h}|_{h=0}=0 \, ,
\ee
so that the previous equation seems to be always satisfied.

However the we must be careful if most of solutions have a very small $\det (M)$. In this case if we 
first sent $N$ to infinity and after we do the derivative with respect to $h$ we can find that the set 
of solutions in not a continuos functions of $h$. Solutions may bifurcate or disappear for any arbitrary 
small variation of $h$ and the previous relations are no more valid. 

This phenomenon has been studied in the framework of infinite range random matrices, where the
function that plays the role of $H$ is the free energy as function of the magnetizations (i.e. the
TAP free energy) in spin glass type models.  One finds that depending on the parameters there are
two phase, one where the supersymmetry is exact, the other where the supersymmetry is spontaneously
broken.  This last phenomenon has been discovered last year and at the present moment one is trying
to fully understand its consequences.

In the framework of Euclidean random theory there are two questions that are quite relevant and may 
be the most interesting  for 
Euclidean Random theory:
\begin{itemize}
	 \item How to construct and to use in a practical way a formalism where  the 
	 supersymmetry  of the problem plays a crucial role?
	 \item How to find out if there is a phase where supersymmetry is spontaneously broken and which are the physical
	 effects of such a breaking.
	\end{itemize}
It is quite likely the response to these questions would be very important.

\section*{Acknowledgements}
I have the pleasure to thank all the people that have worked with me in the subject of random matrices 
and related problems, i.e. Alessia Annibale, Andrea Cavagna, Andrea Crisanti, Stefano Ciliberti, Barbara 
Coluzzi, Irene Giardina, Tomas Grigera, Luca Leuzzi, V\'{\i}ctor Mart\'{\i}n-Mayor, Marc M\'ezard, Tommaso Rizzo, 
Elisa Trevigne,
Paolo Verrocchio and Tony Zee.


\begin{thebibliography}{99}

\bi{MOLTE} T.\ M.\ Wun and R.\ F.\ Loring, J. Chem.  Phys {\bf 97,} 8368 (1992); Y.\ Wan 
and R.\ Stratt, J. Chem.  Phys {\bf 100,} 5123 (1994);  A.\ Cavagna et al., Phys.  Rev.  Lett.  {\bf 83,} 108 (1999); G.\ 
Biroli, R.\ Monasson, J.\ Phys.  A: Math.  Gen.  {\bf 32,} L255 (1999).


\bi{MPZ} M. M\'ezard, G. Parisi, and A. Zee, Nucl.\ Phys.\ {\bf B559,} 689 (1999), 
\bi{MPg} M. M\'ezard and G.Parisi, Phys. Rev. Lett. {\bf 82} (1999)747.

\bi{CoPaVe} B.Coluzzi, G.Parisi and P.Verrocchio, {\em J. Chem. Phys.}
{\bf 112}, 2933 (2000), B.Coluzzi, G.Parisi and P. Verrocchio,
{\em Phys. Rev. Lett.} {\bf 84},306 (2000)

\bibitem{Zee2} A.\ Zee and I.\ Affleck, J.\ Phys: Condens.\ Matter {\bf 12},
8863 (2000).

\bi{V1} V.\ Mart\'\i{}n-Mayor, M. M\'ezard, G. 
Parisi, and P. Verrocchio, ``The dynamical structure factor in topologically
disordered systems'' cond-mat 0008472v1 (unpublished).

\bi{LONG} V.\ Mart\'\i{}n-Mayor, M. M\'ezard, G. 
Parisi, and P. Verrocchio, J.\ Chem.\ Phys.\ {\bf 114}, 8068 (2001).


\bibitem{Proceedings} T.\ S.\ Grigera et al., cond-mat/0104433 (to be
published in Philos. Mag. {\bf B}).

\bibitem{INPREPARATION} T.\ S.\ Grigera et al., to be published.


\bibitem{BOSONPEAK} T.S. Grigera, V. Mart\'{\i}n-Mayor, G. Parisi and
P. Verrocchio, preprint cond-mat/0110129.

\bibitem{Grigera01} 
 T.S. Grigera, V. Mart\'{\i}n-Mayor, G. Parisi and
P. Verrocchio, Phys. Rev. Lett.
{\bf 87} 085502 (2001)


\bi{BOSE}T.S. Grigera, V. Mart\'{\i}n-Mayor, G. Parisi and P. Verrocchio, Phys.  Rev.  
Lett.  {\bf 87} 085502 (2001), T.S. Grigera, V. Mart\'{\i}n-Mayor, G. Parisi and P. 
Verrocchio, cond-mat/0104433 and cond-mat/0301103 (to be published on Nature).


\bibitem{CGP1} A. Cavagna, I. Giardina, G. Parisi, J. Phys. A 
{\bf 30} (1997) 7021.

\bibitem{mod} Cavagna A, Giardina I and Parisi G  1998 {\it Phys. Rev. B} {\bf 57} 11251.  i

\bi{LOC} S. Ciliberti, T.S. Grigera, V. Mart\'{\i}n-Mayor, G. Parisi and P. Verrocchio,

\bi{loc_rev} See Efetof's lectures at thi
\bibitem{TDYN} W. G\"otze and L. Sjogren, , Rep. Prog. Phys. {\bf 55}
241 (1992); W. Kob and H. C. Andersen, Phys. Rev. Lett. {\bf 73}, 1376
(1994).

\bi{PARISILOC} G. Parisi J. Phys A 1{\bf 14} (1981) 735, {\it An Introduction to the Statistical
Mechanics of Amorphous Systems}, in Recent Advances in Field Theory and Statistical Mechanics,
edited by J.-B. Zuber and R. Stora (North-Holland, Amsterdam, Netherlands, 1984).

\bi{Ell} See S.R.
Elliot {\em Physics of amorphous materials}, Longman (England 1983).

\bibitem{Zim} See {\em e.g.} J.\ M.\ Ziman, {\em Models of disorder},
Cambridge University Press, Cambdrige (1979).

\bi{KEYES} T.\ Keyes, J.\ Phys.\ Chem.\ A {\bf 101,} 2921 (1997).

\bi{C}A. Cavagna, Europhys. Lett. {\bf 53}, 490 (2001).

\bi{ab}L. Angelani, R. Di Leonardo, G. Ruocco, A. Scala, and F. Sciortino, Phys.  Rev.  Lett.  {\bf 
85}, 5356 (2000), K. Broderix, K. K. Bhattacharya, A. Cavagna, A. Zippelius, and I. Giardina, Phys.  
Rev.  Lett.  {\bf 85}, 5360 (2000), T.S. Grigera, A. Cavagna, I.Giardina and G. Parisi, Phys.  Rev.  
Lett.  {\bf 88}, 055502 (2002).

\bi{P12} G. Parisi {\it On the origine of the Boson peak}, cond-mat/0301284, Proceeding of the Pisa
conference September 2002, to be published on Journal of Physics; cond-mat/030128 {\it Euclidean
random matrices, the glass transition and the Boson peak}, proceeding of the Messina conference in
honour of Gene Stanley, Physica A in press.


\bi{CaGiPa} A.Cavagna, I.Giardina, G.Parisi, {\em Phys. Rev. Lett.} {\bf 83},
108 (1999)

\bibitem{RFIM} G. Parisi and N. Sourlas, Phys. Rev. Lett. {\bf 43} (1979) 744; 
Nucl. Phys. B {\bf 206} (1982) 321. 
J. Cardy, Phys. Lett. {\bf 125B} (1983) 470.
A. Klein and J. Fernando-Perez, Phys.Lett. {\bf 125B} (1983) 473.

\bibitem{ps2} Parisi G and Sourlas N 1982 {\it Nucl. Phys. B} {\bf 206}, 321.

\bi{POLIA}A Poliakiov, Soviet Phys. JEPT {\bf 50}, 353 (1970).

\bi{MIG} A.Migdal, (private comunication).

\bi{Abou}R. Abou-Chacra, P.W. Anderson, and  D.J. Thouless,
J. Phys. C {\bf 5}, 1734 (1973).
H. Kunz, J. Physique {\bf 44}, L411 (1883).
E.N. Economou and M.H. Cohen, Phys. Rev. B {\bf 5},
2931 (1972).
A.D. Mirlin and Y.V. Fyodorov, J. Phys. A:
Math. Gen. {\bf 24}, 2273 (1991). A.D. Mirlin and Y.V. Fyodorov,
Nucl. Phys. B {\bf 366}, 507 (1991).
 

\bibitem{diluite} P. Cizeau and J.P. Bouchaud, Phys.  Rev.  E {\bf 50}, 1810 (1994).  G.J. Rodgers and A.J.
Bray, Phys.  Rev.  B {\bf 37}, 3557 (1988).  A.J. Bray and G.J. Rodgers, Phys.  Rev.  B {\bf 38}, 11461
(1988).  G. Biroli and R. Monasson, e-print cond-mat/9902032 (1999).
 

\bi{CGP} A. Cavagna, I. Giardina and G. Parisi,
Phys. Rev. Lett. {\bf 83} 108 (1999).

\bi{AZ} A. Zee, I. Affleck,
J. Phys.: Cond. Matter {\bf 12}, 8863 (2000); 


\bi{BOSEREV}Pilla, O. {\em et al.,} Nature of the Short
Wavelength Excitations in Vitreous Silica: An X-Ray Brillouin
Scattering Study. {\em Phys. Rev. Lett.} {\bf 85,} 2136--2139 (2000).
\bibitem{Ruo2} G.\ Ruocco et al., Phys.\ Rev.\ Lett. {\bf 84,} 5788 (2000).

\bibitem{HoKoBi} J.\ Horbach et al., J.\ Phys.\ Chem.\ B
{\bf 103,} 4104 (1999).


\bi{BeLa} S.B. Bembenek and S.D. Laird, {\em J. Chem. Phys} {\bf 104}. 5199 (1996)

\bi{ScDiGa} W. Schirmacher, G. Diezemann and C. Ganter,
{\em Phys. Rev. Lett.} {\bf 81}, 136 (1998)

\bi{Montagna} M. Montagna {\em et al.} {\em Phys. Rev. lett.} {\bf 83}, 
3450 (1999).

\bibitem{MaPaVe} V.\ Mart\'\i{}n-Mayor et al., Phys.\ Rev.\ E {\bf
62,} 2373 (2000).

\bibitem{MOMENTI}
C.Benoit, E.Royer and G.Poussigue, {\em J. Phys
Condens. Matter} {\bf 4}. 3125 (1992), and references therein;
C.Benoit, {\em J. Phys Condens. Matter} {\bf 1}. 335 (1989), G.Viliani
et al. {\em Phys. Rev. B} {\bf 52}, 3346 (1995).
{\sl Phys. Rev. E}.
\bibitem{TRUNCAZIONE} 
P. Turchi, F. Ducastelle and G. Treglia, J. Phys. C {\bf 15}, 2891 (1982).

\bibitem{GoMa} W.\ G\"{o}tze and M.\ R.\ Mayr, Phys.\ Rev.\ E {\bf
61,} 587 (2000).

\bibitem{MPV} M\'ezard M,  Parisi G and Virasoro M.A., {\it Spin glass theory and beyond'', World Scientific} (1987).

\bi{ParisiAndalo} G. Parisi {\it A pedagogical introduction to the replica method for fragile
glasses}, cond-mat/9905318, Phil.  Mag. in press.

\bibitem{juanpe} Cavagna A, Garrahan J P and Giardina I 1998 {\it J. Phys. A: Math. Gen.} {\bf 32} 711.

\bibitem{zinnjustin} Zinn-Justin J, 1989 {\em Quantum Field Theory and Critical Phenomena}, (Clarendon Press, Oxford).
\bibitem{jorge1} Kurchan J 1991 {\it J. Phys. A: Math. Gen.} {\bf 24} 4969.
\bibitem{jorge2} Kurchan J 2002 {\it Preprint cond-mat/0209399}.
\bibitem{discuss} For an illuminating discussion of the problem of removing the 
determinant  in the supersymmetric formalism see \cite{jorge1,jorge2} and \cite{ps2}.

\bibitem{commentBM} A. Bray and M. Moore, e-print {\sf cond-mat/0305620}.

\bibitem{BRST} C. Becchi, A. Stora and R. Rouet, Commun. Math. Phys. {\bf 42} (1975) 127. \\
I.V.  Tyutin {\em Lebdev preprint} FIAN {\bf 39} (1975) unpublished.

\bibitem{CGPM} A. Cavagna, I. Giardina, G. Parisi and M. Mezard, J. Phys. A 
{\bf 36} (2003) 1175.

\bibitem{TAP}D. J. Thouless, P. W. Anderson and R. G. Palmer, Phil. Mag.
{\bf 35} (1977) 593.

\bibitem{CLPR} A. Crisanti, L. Leuzzi, G. Parisi, T. Rizzo, Phys.  Rev.  B, 68 (2003) 174401;
Phys. Rev. Lett. 92, 127203 (2004), Phys. Rev. B 70, 064423 (2004).

\bibitem{CGP3} A. Crisanti, L. Leuzzi, T. Rizzo, Eur.  Phys.  J. B, 36 (2003) 129-136;{ \it
Complexity in Mean-Field Spin-Glass Models: Ising $p$-spin}, cond-mat/0406649.

\bibitem{CGP4} A. Crisanti, L. Leuzzi, Phys.  Rev.  B 70, 014409 (2004), {The spherical $2+p$
spin glass model: an exactly solvable model for glass to spin-glass transition} cond-mat/0407129.

\bibitem{CGP5} A, Annibale, A. Cavagna, I. Giardina, G. Parisi, Phys.  Rev.  E 68,
061103 (2003); A, Annibale, A. Cavagna, I. Giardina, G. Parisi, Elisa Trevigne J.
Phys.  A 36, 10937 (2003); A. Cavagna, I. Giardina, G. Parisi,Phys.  Rev.  Lett.  92,
120603 (2004); A. Annibale, G. Gualdi, A. Cavagna J. Phys.  A: Math.  Gen.  37 (2004)
11311, A. Cavagna,  I. Giardina,  G. Parisi, Phys. Rev. B 71 (2005) 024422.

\bibitem{CR} G. Parisi, T. Rizzo {\it On Supersymmetry Breaking in the Computation of the Complexity
} cond-mat/0401509; {\it Zero-Temperature Limit of the SUSY-breaking Complexity in Diluted Spin-Glass
Models} cond-mat/0411732

\bi{R}T. Rizzo {\it Tap Complexity, the Cavity Method and Supersymmetry}
cond-mat/0403261, T. Rizzo {\it On the Complexity of the Bethe Lattice Spin Glass} cond-mat/0404729;







\end{thebibliography}
\end{document}